\documentclass[12pt]{iopart}          

\usepackage{iopams} 
\usepackage{graphicx}
\usepackage{longtable}
\usepackage{array}
\usepackage{dcolumn}
\usepackage{bm}
\usepackage{float}
\usepackage[british]{babel}
\usepackage{multirow}
\usepackage{caption}
\usepackage{siunitx}

\usepackage{lineno}

\usepackage{rotating}
\usepackage{color}
\usepackage{hyperref}
\hypersetup
{
pdftitle="Cyclotron Radiation Emission Spectroscopy Signal Classification with Machine Learning in Project 8",
pdfauthor="Luis Saldana",
pdftoolbar=true, 
pdfmenubar=true, 
pdfstartview={FitH},
colorlinks=true, 
pdfpagemode=UseNone, 
pdfpagelayout=SinglePage, 
pdffitwindow=true, 
linkcolor=blue,
citecolor=blue,
urlcolor=blue 
}

\usepackage[utf8x]{inputenc}
\usepackage[T1]{fontenc}

\usepackage{footmisc}
\usepackage[font=small]{caption}

\begin{document}

\title[CRES Signal Classification with Machine Learning in Project 8]{Cyclotron Radiation Emission Spectroscopy Signal Classification with Machine Learning in Project 8}

\author{
A.~Ashtari Esfahani$^1$,
S.~B\"oser$^2$,
N.~Buzinsky$^3$,
R.~Cervantes$^1$,
C.~Claessens$^2$,
L.~de~Viveiros$^4$,
M.~Fertl$^{1,2}$,
J.~A.~Formaggio$^3$,
L.~Gladstone$^5$,
M.~Guigue$^6$\footnote{Present address: Sorbonne Universit\'e and Laboratoire de Physique Nucl\'eaire et des Hautes \'Energies, CNRS/IN2P3, 75005 Paris, France},
K.~M.~Heeger$^7$,
J.~Johnston$^3$,
A.~M.~Jones$^6$,
K.~Kazkaz$^8$,
B.~H.~LaRoque$^6$,
A.~Lindman$^2$,
E.~Machado$^1$,
B.~Monreal$^5$,
E.~C.~Morrison$^6$,
J.~A.~Nikkel$^7$,
E.~Novitski$^1$,
N.~S.~Oblath$^6$,
W.~Pettus$^1$,
R.~G.~H.~Robertson$^1$,
G.~Rybka$^1$,
L.~Salda\~na$^{7\mathsection}$,
V.~Sibille$^3$,
M.~Schram$^6$,
P.~L.~Slocum$^7$,
Y.-H.~Sun$^5$,
T.~Th\"ummler$^9$,
B.~A.~VanDevender$^6$,
T.~E.~Weiss$^3$,
T.~Wendler$^4$,
E.~Zayas$^{3\mathsection\mathsection}$
}

\ead{$^{\mathsection}$luis.saldana@yale.edu, $^{\mathsection\mathsection}$ezayas@mit.edu}

\address{$^1$ Center for Experimental Nuclear Physics and Astrophysics and Department of Physics, University of Washington, Seattle, WA 98195, USA}
\address{$^2$ Institut f\"ur Physik, Johannes Gutenberg-Universit\"at Mainz, 55128 Mainz, Germany}
\address{$^3$ Laboratory for Nuclear Science, Massachusetts Institute of Technology, Cambridge, MA 02139, USA}
\address{$^4$ Department of Physics, Pennsylvania State University, State College, PA 16801, USA}
\address{$^5$ Department of Physics, Case Western Reserve University, Cleveland, OH 44106, USA} 
\address{$^6$ Pacific Northwest National Laboratory, Richland, WA 99354, USA}
\address{$^7$ Wright Laboratory, Department of Physics, Yale University, New Haven, CT 06520, USA}
\address{$^8$ Lawrence Livermore National Laboratory, Livermore, CA 94550, USA}
\address{$^9$ Institut f\"ur Kernphysik, Karlsruher Institut f\"ur Technologie, 76021 Karlsruhe, Germany}

\begin{abstract}{
The Cyclotron Radiation Emission Spectroscopy (CRES) technique pioneered by Project 8 measures electromagnetic radiation from individual electrons gyrating in a background magnetic field to construct a highly precise energy spectrum for beta decay studies and other applications. The detector, magnetic trap geometry and electron dynamics give rise to a multitude of complex electron signal structures which carry information about distinguishing physical traits. With machine learning models, we develop a scheme based on these traits to analyze and classify CRES signals. Proper understanding and use of these traits will be instrumental to improve cyclotron frequency reconstruction and boost the potential of Project 8 to achieve world-leading sensitivity on the tritium endpoint measurement in the future.
}\end{abstract}



\section{Introduction}

\begin{figure}[ht!]
\begin{center}
\includegraphics[width=0.7\textwidth]{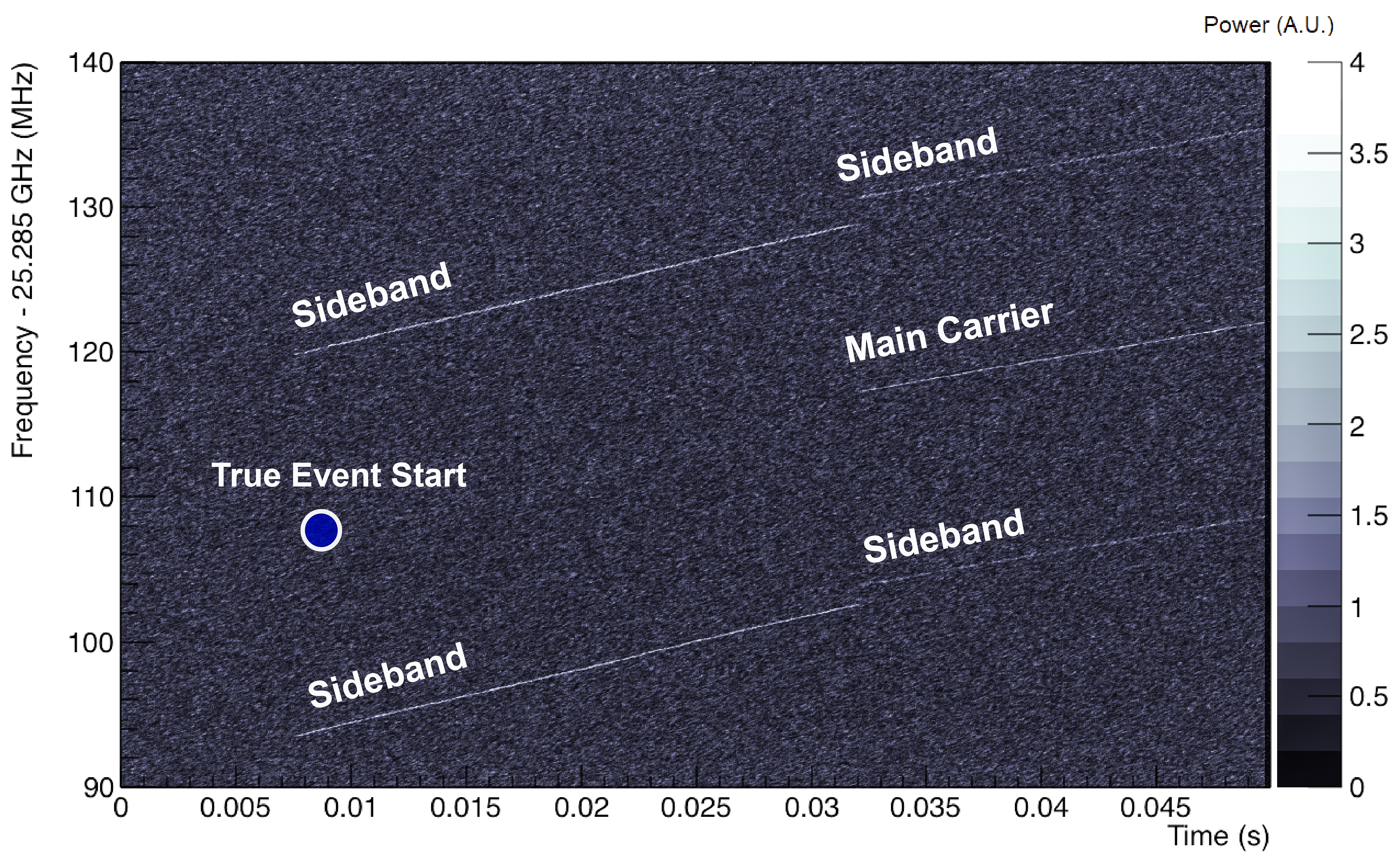}
\caption{A multi-track electron event featuring five tracks. The electron is born in the trap around 7 ms and scatters with a residual gas molecule around 32 ms, abruptly changing the frequency of all tracks. This event is shown for illustrative purposes only and is not from the data sets used in this work.\label{fig:phaseII_event}}
\end{center}
\end{figure}

The Project 8 experiment aims to perform an ultra-precise measurement of the tritium beta decay endpoint to directly measure or constrain the effective mass of the electron anti-neutrino, and to determine the mass hierarchy ordering. To this end, the collaboration has pioneered the Cyclotron Radiation Emission Spectroscopy (CRES) technique \cite{Asner:2014cwa}, in which electromagnetic radiation from the cyclotron motion of individual electrons in a magnetic field $\vec{B}$ is used to reconstruct an energy spectrum from the angular frequency:
\begin{equation}\label{eq:cyclotronfreq}
\Omega_c(t)=\frac{eB(\vec{r},t)}{m_e+K_e(t)/c^2}
\end{equation}
where $B$ is the magnetic field magnitude, $e$ and $m_e$ are the electron charge magnitude and mass, $K_e(t)$ is the electron's kinetic energy, and $c$ is the speed of light. In the non-relativistic regime, this gives a low energy limit of $2.8\times 10^{10}$ Hz in a 1 T magnetic field. A CRES signal is reconstructed via a series of short-time discrete Fourier transforms (DFTs) to produce a frequency spectrum as a function of time (a spectrogram). Due to radiative energy loss, the signal exhibits a pseudo-linear behavior in this time/frequency plane; Figure \ref{fig:phaseII_event} shows an example spectrogram with several such signals. We refer to these CRES signals as tracks.

For this work, we are primarily concerned with the apparatus and data from Phase I of Project 8; the data is from a campaign performed in 2015 with $^{83\textrm{\tiny{m}}}\textrm{Kr}$ as the electron source gas. The Phase I detector \cite{JPhysG} featured a rectangular waveguide to house the source gas and transport emitted radiation from the source to an antenna, as well as a conductive short acting as a reflector opposite the antenna.

\begin{figure}[ht!]
\centering
\includegraphics[width=6.5cm]{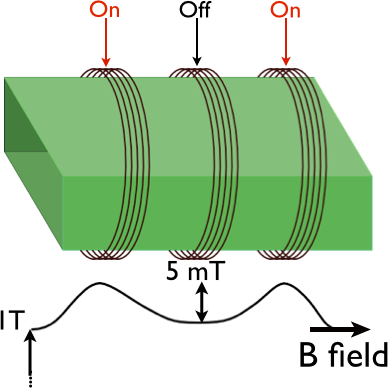}
\caption{Diagram of the magnetic bottle trap in a bathtub configuration. Three coils are wound around the rectangular waveguide whose magnetic fields create a potential barrier along the main axis where a 1 T background field is present. Electrons are constrained to the low-field region between two trapping coils.}
\label{fig:trap}
\end{figure}

The waveguide also sports a configurable magnetic bottle trap formed by pinch coils wound along the axis of the $\sim1$ T background magnetic field, which we call the axial direction or simply $\hat{z}$ by definition. In such a trap, electrons with momentum mostly in the $(\hat{x},\hat{y})$ directions can be constrained in $\hat{z}$ by the small $\mathcal{O}(\textrm{mT})$ influence of the trap coils. In the "bathtub trap" (the only configuration explored here), two coils of equal polarity source the trap by creating a pair of potential barriers for the electrons as illustrated in Figure \ref{fig:trap}. Thus, in addition to cyclotron motion in the $(x,y)$ plane, a trapped electron exhibits a slower axial oscillation $\mathcal{O}(\textrm{MHz})$ as it explores the allowed region in $z$ within the trap. In our experimental setup a full event has an average of approximately 2.5 ms. This axial motion gives rise to a number of rich signal characteristics beyond only the instantaneous cyclotron frequency given by Equation \ref{eq:cyclotronfreq}. In this paper, we will summarize our understanding of these characteristics, the impact on frequency reconstruction, and present a machine-learning (ML) track classification scheme as a first step toward a sophisticated CRES signal analysis. This type of analysis will allow for significant improvement in the energy resolution achievable with CRES, and will be especially beneficial when moving from a monoenergetic krypton source to a continuous tritium source where proper event reconstruction is of paramount importance. Lastly, we study the impact of our ML classification analysis on the extracted tritium endpoint through simulation.

\subsection{Data and Signal Basics}\label{subsec:data-signal-basics}

The $^{83\textrm{\tiny{m}}}\textrm{Kr}$ source emits internal conversion electrons at several energies, which we divide into three nominal groups by the atomic shell of the transition:

\begin{table}[h]
\begin{indented}
\lineup
\item[]\begin{tabular}{@{}*{6}{l}}
\br
\0\0Group&Line&Intensity (total$\equiv 1$)&Energy (eV)&Width (eV)\cr
\mr
17 keV & $K$ & 0.36 & 17830.0 & 2.83 \cr
\multirow{2}{*}{30 keV} & $L_2$ & 0.392 & 30424.4 & 1.84 \cr
& $L_3$ & 0.581 & 30477.2 & 1.4 \cr
\multirow{3}{*}{32 keV} & $M_2$ & 0.067 & 31934.2 & 1.99 \cr
 & $M_3$ & 0.105 & 31941.9 & 1.66 \cr
 & $N_{2+3}$ & 0.016 & 32140.9 & 0.59 \cr
\br
\end{tabular}
\end{indented}
\caption{Relative intensities of $^{83\textrm{\tiny{m}}}\textrm{Kr}$ conversion electron lines under study in this work. From \cite{Picard1992} and \cite{Isotopes}.}
\label{table:kr-lines}
\end{table}

Each transition is monoenergetic, up to a natural linewidth of order eV, which is substantially less than the energy resolution of Phase I. The 17 keV data is closest to the tritium endpoint (18.6 keV), and the higher-energy peaks provide important insight on the relative energy dependence of signal characteristics.

Electron signals emitted in the magnetic trap are received by an antenna and processed by a cryogenic receiver chain described in \cite{Asner:2014cwa}. The signal is down-mixed twice using local oscillators and sampled at 200 MHz by a real-time spectrum analyzer (RSA). The RSA triggers an acquisition when it detects a high Fourier excess, and writes time-domain data for 10 milliseconds per trigger with a pre-trigger time of 1 ms. The acquisition is then processed with a series of DFTs of size 8192 samples (0.04096 ms) to produce a spectrogram like the one in Figure \ref{fig:phaseII_event}, which displays a CRES event. The resulting spectrograms are scanned for high-power bins in collinear groupings called tracks; for an in-depth discussion the track finding algorithms and procedure, we refer to \cite{KofronThesis}. The initial frequency of a track signal is called the start frequency, and it is the start frequency of an electron event that contains primary energy information needed for spectrum reconstruction.

All events considered in this study are subject to a cut on the start time of the first track within $\pm 0.25$ ms of the pre-trigger time. This retains only events which have promptly triggered an acquisition and removes those which only triggered after some time due to sufficiently enough power; these low-power tracks often start even before the acquisition window, which prohibits a measurement of the start frequency altogether. Furthermore, in preparing the training set for the machine learning analysis we consider only the first track(s) in such events. In the example of Figure \ref{fig:phaseII_event}, the second set of tracks would be cut, which is useful in labeling the ground truth, a process described in Section \ref{sec:optimization}; by only considering the first set of tracks in an event we can confidently label separate signal classes using both frequency information and other parameter space cuts.

\section{The Need for Classification}
\label{sec:need_for_classification}
In this section, we summarize our understanding of relativistic electron dynamics in Project 8 traps to motivate the classification scheme and elucidate the resultant properties of reconstructed CRES signals. An analysis that properly extracts and uses the information in these signal properties is key to obtaining a precise, well-understood energy spectrum.

\subsection{Axial Motion Considerations} \label{subsec:cres_signal_waveguide}

We parameterize the axial motion of an electron with the pitch angle $\theta(t)$, defined as the angle between the momentum vector and the magnetic field:
\begin{equation}
\cos{\theta}(t)=\frac{\vec{p}\cdot \vec{B}(t)}{pB(t)}=\frac{p_z(t)}{p}\ \mathrm{for}\ \vec{B}=B_z \hat{z}
\label{eq:costheta}
\end{equation}
For a trapped electron, $p_z$ has an oscillatory time dependence and therefore so does the instantaneous pitch angle.\footnote{The fractional energy loss over the timescale of the axial period is small, so we may treat the total momentum $p$ as a constant here.} By definition, $\theta=90^\circ$ at the turning points of the axial oscillation as $p_z=0$; at the center of the trap, the pitch angle reaches a minimum along with $B_z$. Thus, the range in $z$ explored by an electron is fully characterized by (a) the trap geometry and (b) the minimum pitch angle. Going forward, we will simply use $\theta$ to refer to this minimum pitch angle, rather than the time-varying instantaneous pitch angle $\theta(t)$. In fact, there is also a limit to the smallest pitch angle value due to conservation of energy which depends on the ratio of the trap depth and maximum magnetic field, for more on this see \cite{PhenomPaper}. At a nominal 4 mT trap depth the lower bound is about 86 degrees and may be higher depending on the chosen geometry.

A detailed mathematical discussion of trapped electron dynamics and the resultant CRES signal characteristics is outside the scope of this work but may be found in \cite{PhenomPaper}; we refer to their nomenclature for the rest of the paper. There, a phenomenological model is developed using approximate Project 8 magnetic trap geometries to analytically describe the motion of electrons with $\theta<90^\circ$. The model includes a short opposite the antenna, as is present in the Phase I detector; the result is an energy (frequency) and position-dependent interference effect between the incident and reflected radiation. The power spectrum $P(\omega)$ is calculated from the Poynting vector in the axial direction, and decomposed as a sum of waveguide modes. For a single mode denoted by $\lambda$, we take advantage of the quasi-periodic motion of the electron in the trap to express the power averaged over the axial period with Equation (45) in \cite{PhenomPaper}, reproduced here:
\begin{eqnarray}\label{eq:power_short}
 P_\lambda(\omega) & = 4 P_{0,\lambda} \sum _{n=-\infty} ^{\infty} \left\vert a_n \left(k_\lambda\right)\right\vert^2 \cos^2\left[(z_t+l)k_\lambda\right]\\
&\times \left[\delta(\omega - k_\lambda v_{p,\lambda})) + \delta(\omega + k_\lambda v_{p,\lambda})\right]\ \mathrm{with}\ k_\lambda = \frac{\Omega_0+n\Omega_a}{v_{p,\lambda}} \nonumber
\end{eqnarray}

\noindent where $P_{0,\lambda}$ and every $a_n$ are amplitudes dependent on the magnetic trap shape, $z_t+l$ is the distance from the short to the trap center, and $v_{p,\lambda}$ is the mode phase velocity. This equation describes a comb-like spectrum with power concentrated at a central frequency $\Omega_0$ (the average cyclotron frequency) and at frequencies shifted by integer multiples $n$ of the axial oscillation frequency $\Omega_a$. In the bathtub trap geometry $\Omega_0=\Omega_c\times F(\tan^{-1}\theta)$ where the function of tangent includes parameters describing the length and depth of the trap. The pitch angle dependence of $P_\lambda(\omega)$ comes from $\Omega_0$, $\Omega_a$, and in turn $k_\lambda$; in particular, the coefficients $a_n(k_\lambda)$ describe the relative strength of each peak in the comb structure. Considerable discussion about these coefficients and their calculation for some models is included in \cite{PhenomPaper}.

We refer back to Figure \ref{fig:phaseII_event} in the previous section, now equipped to understand how this example event illustrates the behavior of Equation \ref{eq:power_short}:

\begin{enumerate}
\item[(a)]The signal takes the form of multiple parallel tracks corresponding to the different values of the frequency band order $n$. We call this structure a multi-peak track (MPT).
\item[(b)]At approximately 32 ms, the electron scatters with a residual gas molecule and changes the makeup of the MPT; this is consistent with an abrupt change in the pitch angle. In particular, the frequency and power of individual tracks is observed to be pitch-angle-dependent as expected.
\item[(c)]By contrast, the track slope $-$ the change in frequency (energy) with respect to time $-$ encodes the total radiated power and thus does not vary within tracks of any one MPT.
\end{enumerate}

\begin{figure}[h!]
\begin{center}
\includegraphics[width=1.0\textwidth]{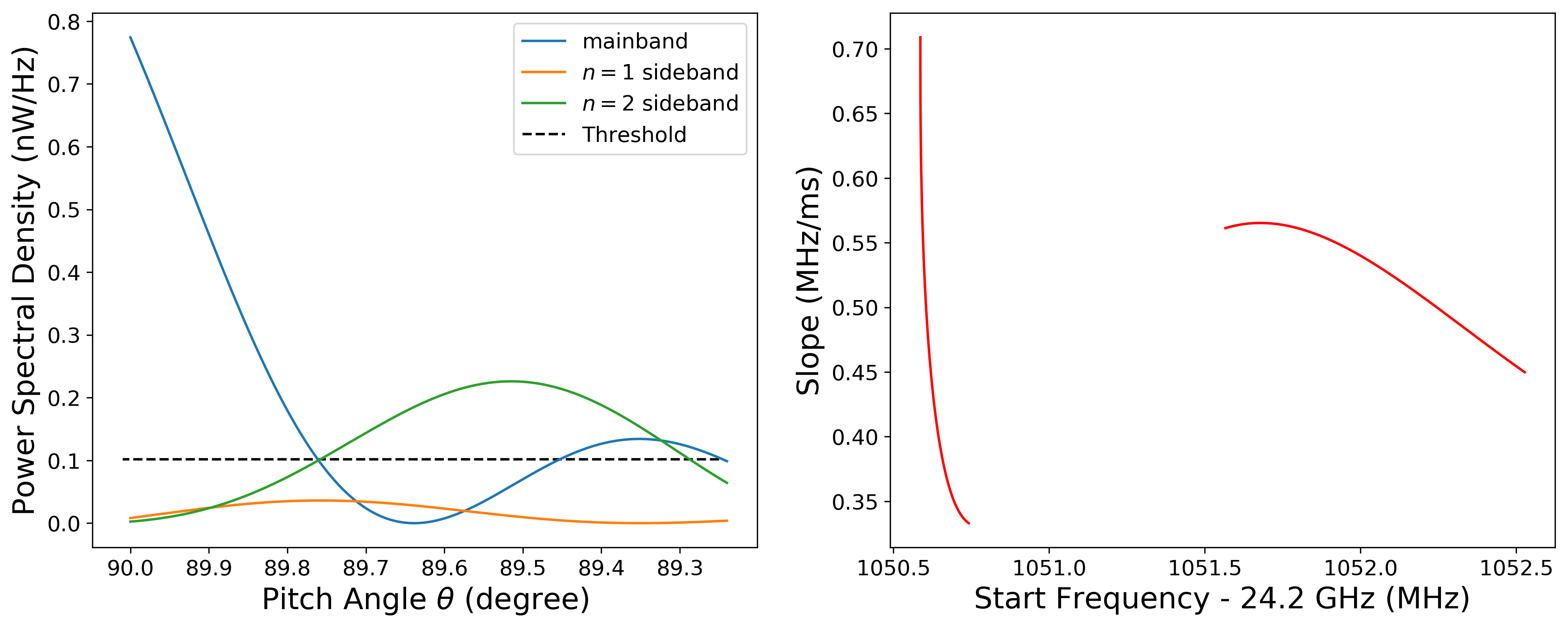}
\caption{Left: Power distribution of a 32 keV electron signal in a bathtub-type Project 8 trap as dependent on pitch angle due to the rectangular waveguide with a short. For a threshold as shown, only the mainband and $2^{\mathrm{nd}}$-order sidebands surpass the detection threshold while the $1^{\mathrm{st}}$-order sideband is suppressed. Right: As a result, the slope of the mainband, which is directly proportional to the total detected power, suffers a discontinuity in frequency. \label{fig:power_vs_theta}}  
\end{center}
\end{figure}

The dependence of the individual band power on the pitch angle for $n \leq 2$ is shown in Figure \ref{fig:power_vs_theta} for a 32 keV electron in a bathtub trap; it is this individual band power, and not the total power, that corresponds to a single track in the spectrogram. Since high-order bands are never powerful enough for reconstruction, we in general restrict our discussions to only the mainband ($n=0$) and one detected sideband order: $n=1$ or $n=2$ depending on the short interference effect. The dashed line in Figure \ref{fig:power_vs_theta} shows an example detection threshold that is met only by the mainband and the $n=2$ sideband for different but partially overlapping ranges of the pitch angle. This creates three allowed track types based on the pitch angle and the band order:

\begin{enumerate}
  \item Mainband high pitch angle: closest to $90^\circ$ for $n=0$.
  \item Mainband low pitch angle: far from $90^\circ$ for $n=0$.
  \item Sidebands: a single range for $n=2$.
\end{enumerate}

Since the short interference is a wavelength-dependent effect, we expect the specific nature of the allowed pitch angle regions to vary with frequency. Indeed, in Project 8 Phase I the 32 keV tracks are well described by Figure \ref{fig:power_vs_theta} and the three cases above; but the story is very different at 17 keV, where the mainband is almost completely suppressed and we detect only the $n=1$ sidebands. For further discussion of this effect where the $n=1$ sideband is visible we refer the reader to Section VII of \cite{PhenomPaper}. This exemplifies the powerful influence of the short, and the importance of understanding sideband effects.

\subsection{Radial Gradient Effects}
\label{sec:gradB}

So far, we have treated the axial motion as independent of the $(x,y)$ plane on the basis that the magnetic field varies only with $z$, i.e. $\nabla \vec{B}$ is always parallel to $\vec{B}$. However, to improve our description of the electron dynamics we must consider a small radial gradient of the form $\nabla \vec{B} \times \vec{B}$\footnote{In fact, the assumption that $\nabla \vec{B} \times \vec{B} = 0$ contradicts the Maxwell equations and thus is clearly unphysical.}. This causes the guiding center of the cyclotron orbit to precess slowly (compared to the axial motion) in $x$ and $y$, which in effect perturbs the one-dimensional magnetic trap profile $B_z(z)$ with a slow time dependence. A small anti-symmetric tipping of the trap coils at angle $\psi$ from $\hat{z}$ is the simplest way to recreate this $\nabla \vec{B} \times \vec{B}$ perturbation in simulation. Consequently, this gradient induces a periodic drift in the axial frequency $\Omega_a$:
\begin{equation}
\Omega_a'(t) \approx \Omega_a\left(1+\frac{r}{l_1}\sin\psi\sin(\Omega_m t)\right) 
\end{equation}

\noindent where $r$ is the radial position of the electron, $l_1$ is the characteristic trap length and $\Omega_m$ is the drift frequency.

\begin{figure}[h!]
\centering
\includegraphics[width=0.60\textwidth]{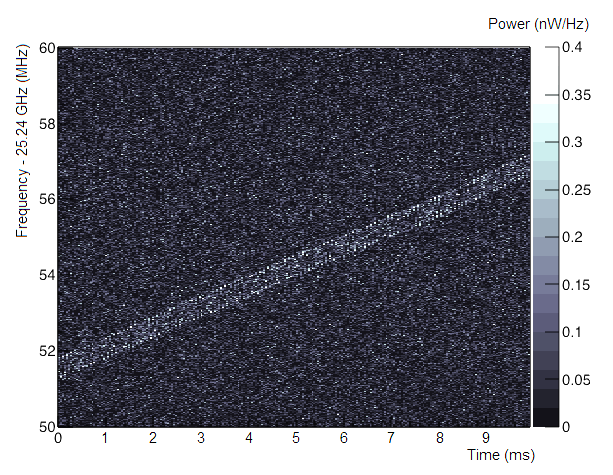}
\caption{Close examination of a spectrogram with a sideband signal. The frequency oscillates due to radial gradients in the magnetic field, depositing power over a range of roughly 500 kHz and most concentrated at the turning points of the oscillation. This oscillation corresponds to the apparent thickness of the track.}
\label{fig:sb_example}
\end{figure}

The effect of this precession on the track signal now becomes clear: if the axial frequency varies sinusoidally, so will the frequency of the $n>0$ (sideband) tracks. This oscillation has been observed in Phase I data, where it manifests as a track with an appreciable "width" in frequency; Figure \ref{fig:sb_example} shows an example sideband track with this quality. Since the observed period of precession is $\sim 100\ \mathrm{\mu s}$, which is comparable to the DFT length ($40.96\ \mathrm{\mu s}$), the oscillation can be seen to some extent directly in the spectrogram.

This observed frequency oscillation represents one of the primary motivations for a machine learning approach to signal classification. Its effect on the spectrogram is clear as illustrated in Figure \ref{fig:sb_example}, and it is a unique property of sideband tracks which demonstrates the power to discriminate from mainbands. By extracting information from the spectrogram around a track, we can then apply machine learning techniques to identify sideband oscillation when it is present and label the track accordingly.

\subsection{Energy Correction}
\label{subsec:energy_corrections}

The cyclotron frequency $\Omega_c$ of a MPT structure can be calculated using the phenomenological model in \cite{PhenomPaper} if both the reconstructed mainband frequency $\Omega_0$ and the pitch angle $\theta$ of the electron are known. The result is in effect an energy correction, where the kinetic energy (and thus from many events, the energy spectrum) is calculated from the true cyclotron frequency rather than the frequency of any one reconstructed track. It is the end goal of the classification scheme to accomplish exactly this; first the identity of the mainband track must be established, and then pitch angle information extracted. We may extract the pitch angle in two ways:

\begin{enumerate}
\item \textit{Axial Frequency}: for MPTs with a mainband and one or more sidebands, the axial frequency $\Omega_a$ is the frequency difference between the mainband and a sideband track divided by its order $n$. $\Omega_a$ may also be determined the same way from an event with multiple sidebands but no mainband. In Phase I, these cases comprise a minority of our data at about $\sim10\%$.
\item \textit{Track Slope}: for MPTs with only a mainband, the pitch angle may be extracted from the track slope, which is proportional to the total radiated power in Equation \ref{eq:power_short}. Such cases comprise the majority of the data used for this work at about $\sim90\%$.
\end{enumerate}

The second method listed above has an ambiguity that must be addressed: in general, the track slope alone does not uniquely determine the pitch angle (this is evident from Figure \ref{fig:power_vs_theta}). To resolve this issue, we must also differentiate between mainbands of the high and low pitch angle regions. Our task is then to assign every track an appropriate topological label from the list in Section \ref{subsec:cres_signal_waveguide}: mainband high pitch angle, mainband low pitch angle, or sideband. Classification into these three groups will allow for an accurate measurement of the mainband frequency, the pitch angle, and in turn the true kinetic energy can be determined from the cyclotron frequency.

We approach this task with a machine learning model that uses a supervised learning method for classification. The overarching goal of the classification program is to use only those track features that are intrinsic to the signal itself and, through their inclusion, have the capability to improve the accuracy and robustness of the signal identification. As Project 8 moves to a tritium source, such a classification scheme will be vital to make an accurate measurement of the continuous spectrum and reach meaningful conclusions about the endpoint. In the remaining sections, we will develop the classification scheme and present the results on krypton Phase I data at all three energies of interest. The next steps of pitch angle calculation and the resultant energy correction are not yet implemented, but are of course a primary focus of future work to realize the full potential of track classification. We will also discuss the future impact of a more-developed classification process in the context of tritium endpoint sensitivity and full event reconstruction.

\section{Signal Analysis and Feature Extraction}
\label{sec:feature_extraction}

The track features which we use for classification result from two separate analysis techniques: primary track finding and the rotate-and-project algorithm. These methods give us a total of 14 parameters that we use in training the ML classification model. In this section, we describe the calculation of these parameters.

\subsection{Primary Track Parameters}
\label{subsec:primary-track-params}

Primary track finding is the process of collecting high-power spectrogram bins into linear track signals, described thoroughly in \cite{KofronThesis}. At its conclusion, several parameters are formally calculated to describe each track candidate; these include the slope, start frequency, time length, and many others. We use the following three quantities as inputs to the classification model:

\begin{itemize}
\item{\texttt{TotalPowerDensity} (W/Hz): the sum of power spectral density values in all bins that comprise the track cluster.}
\item{\texttt{TrackSlope} (Hz/s): the slope of the track as extracted from regression analysis and a Hough transform \cite{Hough}.}
\item{\texttt{TimeLength} (s): the difference between the track end time and start time.}
\end{itemize}

\begin{figure}[h!]
\begin{center}
\includegraphics[width=0.55\textwidth]{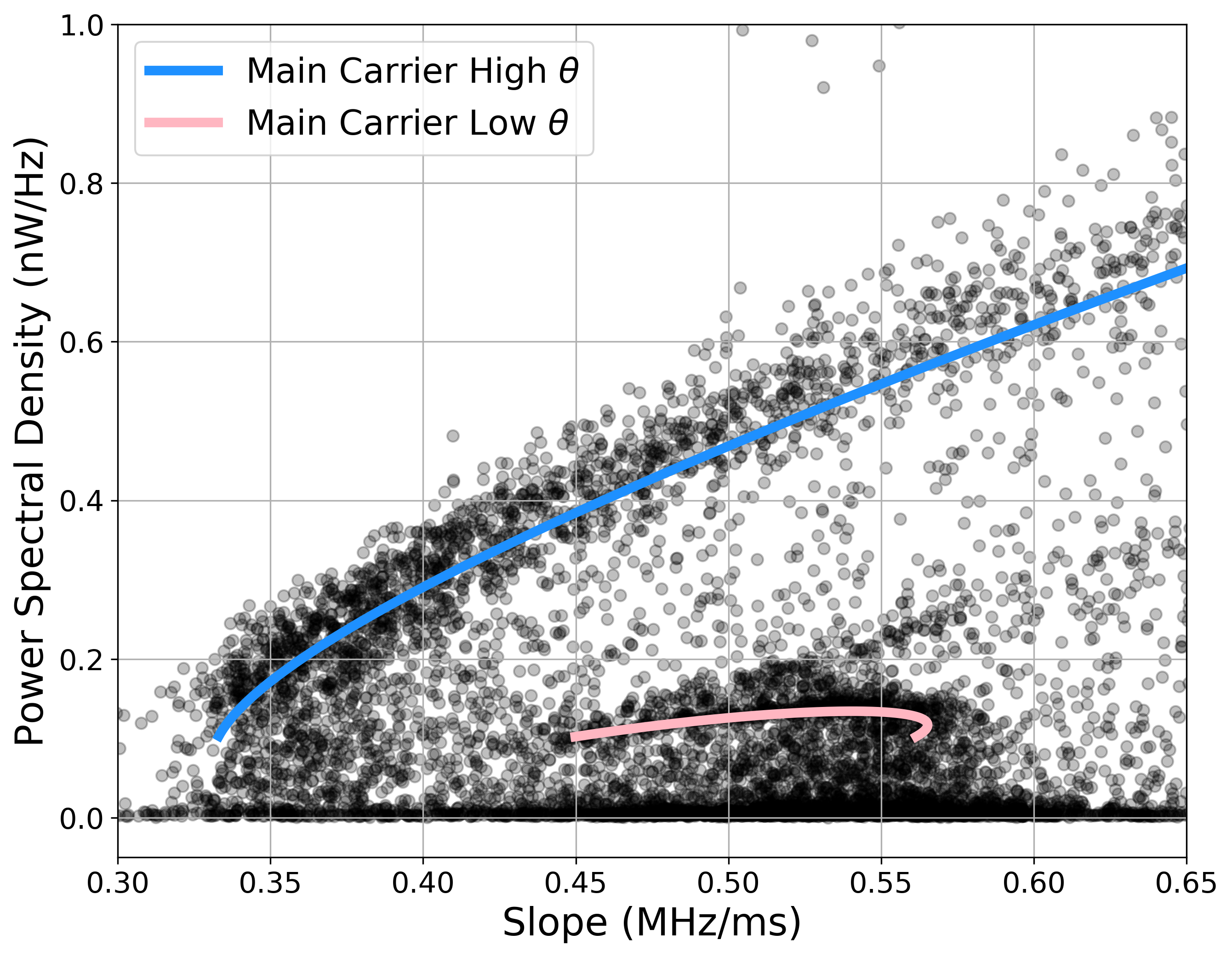}
\caption{Power and slope correlations in Phase I 32 keV bathtub trap tracks (black scatter). The phenomenological model fit is overlaid for high (blue) and low (pink) pitch angle carriers, which demonstrates the well-resolved separation of mainband populations with disjoint pitch angles. \label{fig:main_carrier_difference}} 
\end{center}
\end{figure}

Recall that the slope of a track is directly proportional to the total power emitted by the electron, and the slope and individual track power together determine the pitch angle information as illustrated in Figure \ref{fig:power_vs_theta}. Thus, the correlation between slope and track power has strong discrimination power between regions of high and low pitch angle. Figure \ref{fig:main_carrier_difference} illustrates this correlation for 32 keV electrons and shows a good separation of the two mainband populations, which agrees with our understanding of the physical process from the phenomenological model. In this figure the sideband events populate the low PSD range across all slopes. This provides a clear motivation for the use of these two features as inputs to the classification model.

The utility of the track length is primarily based in its correlation with the track power as well. Mainband tracks in general have a strong profile with power concentrated in one or a few of the frequency bins at each time slice. By contrast, sideband tracks often have power distributed across many bins due to the effect of the radial magnetic field gradient discussed in Section \ref{sec:gradB}. Sidebands also contain less power overall in the case of the 30 and 32 keV peaks. Consequently, the track power can have considerable dependence on the number of points which comprise the track, and the track length  helps to bolster the discrimination ability between all three types in conjunction with the slope and power.

\subsection{Rotate-and-Project Distribution}
\label{subsec:rotate-project}

Radial magnetic field gradients create a sinusoidal variation in the axial frequency, effectively smearing sideband track power over several frequency bins while leaving the mainband track untouched. To extract a set of parameters from the spectrogram that quantify this difference, we first simplify the problem. After primary track finding, spectrograms with a known track are reprocessed with a "Rotate-and-Project" operation, where they are effectively projected along the axis perpendicular to the track. This reduces the analysis to one dimension $-$ the projected spectrum $-$ while preserving the most useful information about the track from the full spectrogram. The precise procedure is as follows:

\begin{enumerate}
\item{The known track is characterized by a slope $q$ and intercept $f_0$.}
\item{The full spectrogram is reduced to a sparse spectrogram of only points that have {SNR > 4.0} and that lie within the time bounds of the track. These points will be described by $(t_j, f_j)$ where $t$ denotes the time coordinate, $f$ the frequency coordinate and $j$ the point index.}
\item{The projected spectrum $s$ at bin $k$ is calculated as a function of the intercept, which we call $\beta_k$ and which sweeps the range $f_0 \pm \Delta f$ in discrete steps of $\delta\beta$ (both $\Delta f$ and $\delta\beta$ are runtime-configurable parameters):
\begin{equation}
s_k = \sum_j \exp{-\frac{(f_j-qt_j-\beta_k)^2}{2\sigma^2}}
\label{rpspect}
\end{equation}

\noindent where $\beta_k=f_0-\Delta f + k\ \delta\beta$, $q$ is the track slope, and $\sigma$ is another runtime-configurable variable which describes the resolution of the spectrum. This calculation is a kernel density estimation with a Gaussian kernel and bandwith $\sigma$.}
\end{enumerate}

At a minimum, $\sigma$ should reflect the inherent uncertainty in each point location, which is roughly the bin size; this way, the spectrum is not strongly affected by how precisely the choices of $\Delta f$, $\delta\beta$, or the reconstructed value $f_0$ coincide with the discrete binning of the spectrogram. $\sigma$ can also be made much larger than the bin size, and the projected spectrum gains sensitivity to structures which span a similarly larger bandwidth; however to retain good sensitivity to the sharp mainband tracks, we keep $\sigma$ similar to the bin width. There is no advantage to matching it exactly with the bin size, so for convenience we choose $\sigma= 50$ kHz which is approximately 2 bins. For the step size, we choose $\delta\beta=25$ kHz, or half the resolution $\sigma$ and approximately 1 bin. The only requirements on the sweep range $2\Delta f$ are that it should be many times larger than the frequency bin size, and at a minimum large enough to capture the full amplitude of the sideband oscillation ($\sim 1$ MHz). We choose $\Delta f=4$ MHz.

\begin{figure}
  \centering
  \begin{tabular}{@{}c@{}}
    \includegraphics[width=0.45\textwidth]{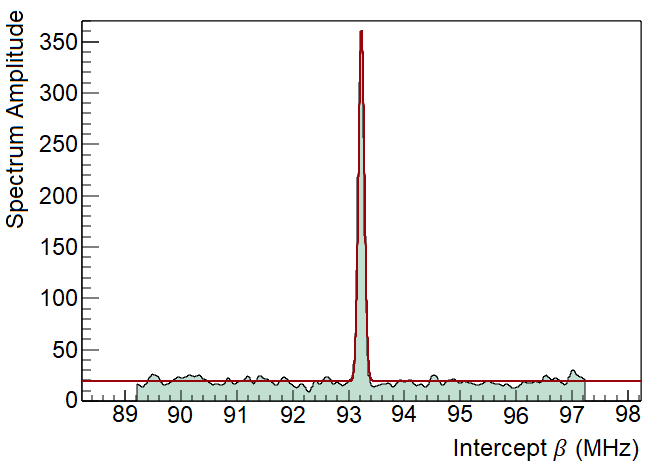} \\[\abovecaptionskip]
    \small (a) Mainband track projected spectrum
  \end{tabular}
  \vspace{\floatsep}
  \begin{tabular}{@{}c@{}}
    \includegraphics[width=0.45\textwidth]{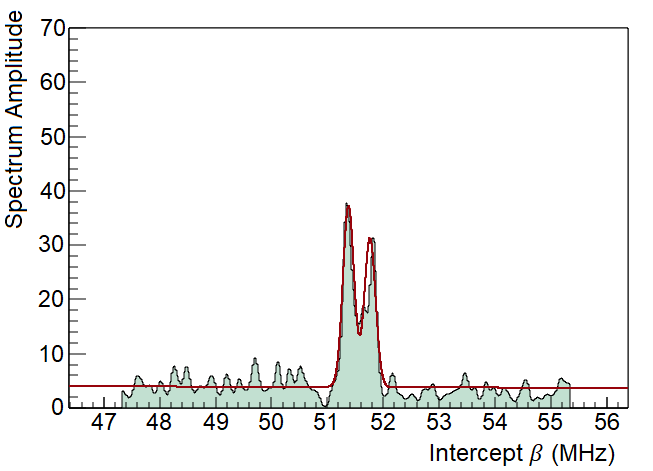} \\[\abovecaptionskip]
    \small (b) Sideband track projected spectrum
  \end{tabular}
  \caption{Spectra from the rotate-and-project analysis for a typical (a) mainband track and (b) sideband track. The mainband track is sharply peaked, whereas the sideband track is spread over a wider frequency range and doubly-peaked. Each spectrum also shows the associated Gaussian fit for comparison. Note the y-axis in (b) is scaled down in comparison to (a). The sideband used for (b) is the same track as illustrated in Figure \ref{fig:sb_example}.} \label{fig:rotate_project_spectra}
\end{figure}

Figure \ref{fig:rotate_project_spectra} shows typical projected spectra corresponding to a mainband and a sideband track. The qualitative differences between the two remain clear: sideband spectra typically feature a wide double peak structure, contrasted with the sharp and high-amplitude profile of the mainband spectrum. The sideband spectrum amplitude is largest near the edges of the signal region because the axial motion is slowest at its turning points, thus depositing more power per bin; this effect can also be seen in the spectrogram (Figure \ref{fig:sb_example}). Next, we use the ROOT library TSpectrum \cite{ROOT} to characterize peaks in the projected spectrum. This library fits a linear background $b_k=ak+b$ and labels a point $k$ as a peak if it meets all of the following criteria:

\begin{enumerate}
\item{Value is at least twice that of the background level: $s_k \geq 2\ b_k$.}
\item{Peak amplitude meets or exceeds a minimum fraction $r$ of the highest peak: $s_k - b_k \geq r \ \sup_j\{ s_j - b_j \}$.}
\item{Value is a local maximum within $m$ bins: $s_k~>~s_j\ \forall\ j\ \mid\ 0 < |j-k| \leq m$. The frequency range corresponding to $m$ bins is $m\ \delta\beta$.}
\end{enumerate}

We choose $r=0.4$ and $m=5$. The values of these and the other configurable parameters from Equation \ref{rpspect} are listed in Table \ref{table:rotate-project}. Once the peak locations are determined, the full spectrum is fit to a sum of $n$ Gaussian functions where $n$ is the number of peaks found. Only a handful of tracks in our studies produced a spectrum with 3 or more peaks, and none with more than 6.

\begin{table}[h]
\begin{indented}
\lineup
\item[]\begin{tabular}{@{}*{2}{l}}
\br
\0\0Parameter&Value\cr
\mr
$\Delta f$ & \04 MHz \cr
$\delta\beta$ & 25 kHz \cr
$\sigma$ & 50 kHz \cr
$r$ & \00.4 \cr
$m$ & \05 \cr
\br
\end{tabular}
\end{indented}
\caption{Standard parameter values for rotate-and-project analysis.}
\label{table:rotate-project}
\end{table}

From the spectrum $s_k$ and the results of the Gaussian fit, we extract a total of 11 additional parameters for use with the classifier:

\begin{itemize}
\item{\texttt{Average}, \texttt{RMS}, \texttt{Skewness}, \texttt{Kurtosis}: First four statistical moments of the spectrum $s_k$. \texttt{Average} and \texttt{RMS} are in units of MHz, and \texttt{Average} is shifted by $f_0$ so that 0 corresponds to the center of the spectrum.}
\item{\texttt{MeanCentral}, \texttt{SigmaCentral}, \texttt{NormCentral}, \texttt{MaximumCentral}: Extracted fit parameters of the Gaussian with mean closest to $f_0$ (the most central peak). \texttt{MeanCentral} is shifted by $f_0$ as described above, and \texttt{MaximumCentral} = $b_0 + (2\pi)^{-1}$ \texttt{NormCentral} / \texttt{SigmaCentral} where $b_0$ is the background level at the peak location. Both \texttt{MeanCentral} and \texttt{SigmaCentral} are in units of MHz.}
\item{\texttt{NPeaks}: Number of peaks found by TSpectrum for the Gaussian fit.}
\item{\texttt{RMSAwayFromCentral}: RMS $\left(\langle s^2 \rangle - \langle s \rangle^2\right)^{1/2}$ of points greater than 3 times \texttt{SigmaCentral} away from the most central peak.}
\item{\texttt{CentralPowerFraction}: average value $\langle s \rangle$ of bins within 3 times \texttt{SigmaCentral} of the most central peak divided by the average value of all points in the spectrum.}
\end{itemize}

In the event that the Gaussian fit fails to converge, all of the parameters that depend on it are obviously unreliable and some can be undefined. To circumvent this, we simply remove any tracks from the analysis that have an unsuccessful fit, or which have any parameters undefined as a result of an improper fit. These represent between 5$-$10\% of all tracks in various data sets we have used. Combining these with the track parameters discussed in the previous subsection, we have a total of 14 parameters to use for machine-learning-based classification. The slope, power, and track length have the ability to distinguish between all three track topologies based on the phenomenological model; the projected spectrum provides an additional 11 parameters which distinguish sidebands from the two mainband topologies. In the next section we discuss the implementation of this new analysis.

\section{Supervised Classification}\label{sec:optimization}

We take a machine learning approach towards signal classification across 17, 30, and 32 keV data sets (see Section \ref{subsec:data-signal-basics}) using a Support Vector Machine (SVM) \cite{Cortes1995} classifier optimized via supervised learning. The result of training the SVM is a nominal decision function that takes data points (track parameters) as 14-dimensional vector inputs and predicts a class label with a given accuracy and precision. In this section we briefly discuss the overall training scheme with details left to Appendix A.

To obtain the training, cross-validation, and test sets over which the classifier is optimized we make a series of parameter-space cuts in the data. The first cut is on the start time as discussed in section \ref{subsec:data-signal-basics}, which selects only tracks that promptly trigger the RSA acquisition. From these, we assign ground-truth labels using two independent fits to the phenomenological model: first, we fit the predicted behavior of slope with respect to the start frequency with the energy assumed to be known exactly. Points within a fixed Euclidean distance in the slope/frequency space of the model predictions were labelled as main carriers, either high-$\theta$ or low-$\theta$ accordingly. Second, we label the remaining sideband tracks using the relative track power with respect to frequency, again with the energy fixed. Tracks outside the inclusion regions for every label are simply discarded. This yields a total of 7,347 tracks for optimization. Labeled tracks are further split in a $67\%/33\%$ fashion for training (training and cross-validation together) and testing, respectively. In performing the split, we keep the relative ratios of classes in mind so as not to introduce biases during training.

It is worth noting briefly that although a ground-truth labelling informed by a proper simulation is likely more desirable, our understanding of these pitch angle effects with the phenomenological model was very new at the time, and our simulation tools were not equipped to incorporate them easily. We have high confidence in the accuracy of the training set labels as described here, so the concern is a minor one. Furthermore, the method described above for labeling is only suitable for training purposes and not for classification since, in generating the fits to the selected subset of data, we assume that the energy is known exactly (with a small margin of error represented by a Euclidean distance in the respective parameter space). In the final data, which we wish to classify, this assumption is not valid for all the tracks we reconstruct. 

The implementation of the classifier is performed with the python-based ML library Scikit-learn \cite{scikit-learn}. In Scikit-learn, template SVMs are implemented by a Cython wrapper around the powerful library LIBSVM \cite{libsvm}. In training the SVM classifier we in parallel optimize the model's hyperparameters $C$ and $\gamma$ which may influence bias and overfitting. $C$ encodes the leniency of the SVM in trading misclassification for model stability (influencing overfitting) and $\gamma$ dictates the influence of training points defining the decision boundary to the rest (influencing bias and variance). We then test the competency of the optimized model on the test set and use the accuracy and the Area Under the Receiver Operating Characteristic (AUROC) as performance metrics. For our multi-class application (two mainband classes and one sideband class), we average the individual Receiver Operating Characteristic (ROC) curves to report the overall AUROC metric.

\section{Results}

Here we report the results of the track classifier as trained on different combinations of the $^{83\textrm{\tiny{m}}}\textrm{Kr}$ line groupings discussed in Section \ref{sec:optimization}. We will show that the optimized SVM classifier can distinguish the three different track topologies with great accuracy and robustness, allowing us to obtain clean spectra across all energy ranges.

\subsection{Narrowband Classifier}\label{subsec:narrowband_results}

In the case of a tritium spectrum, the region of interest will be a window spanning approximately 4 keV (200 MHz) around the endpoint value $Q=18.6\textrm{ keV}$. A classifier will be necessary to understand the CRES signal in this region if sidebands are present and, overall, if energy corrections are to be applied in an event-by-event fashion. With $^{83\textrm{\tiny{m}}}\textrm{Kr}$ as a calibration source gas, we may first study the classifier results by training our model on the 30 keV peaks and applying it to both 30 and 32 keV peaks simultaneously. This 2 keV energy separation\footnote{Including upper and lower sidebands the energy range is wider, at approximately 3.25 keV.} serves as a test of classifier reliability across an energy range similar to the tritium window; we call this configuration the narrowband classifier.

The results of the optimization outlined in the previous section (in detail in Appendix A), picked from a cross-validation accuracy of $92.0\pm0.8\%$ as a mean over 3-folds, gives us the following SVM hyperparameters:

\begin{eqnarray*}
C &= 108.01 \\
\gamma &= 2.947\times 10^{-3}.
\end{eqnarray*}
These values are indicative of a "smooth" model ($\gamma \ll 1$) which captures little of the data complexity in the feature space, but is balanced by the large value of $C \gg 1$ which allows for highly nonlinear terms in the loss metric minimization, recovering some complexity in the decision plane.

\begin{figure}
\begin{center}
\includegraphics[width=0.60\textwidth]{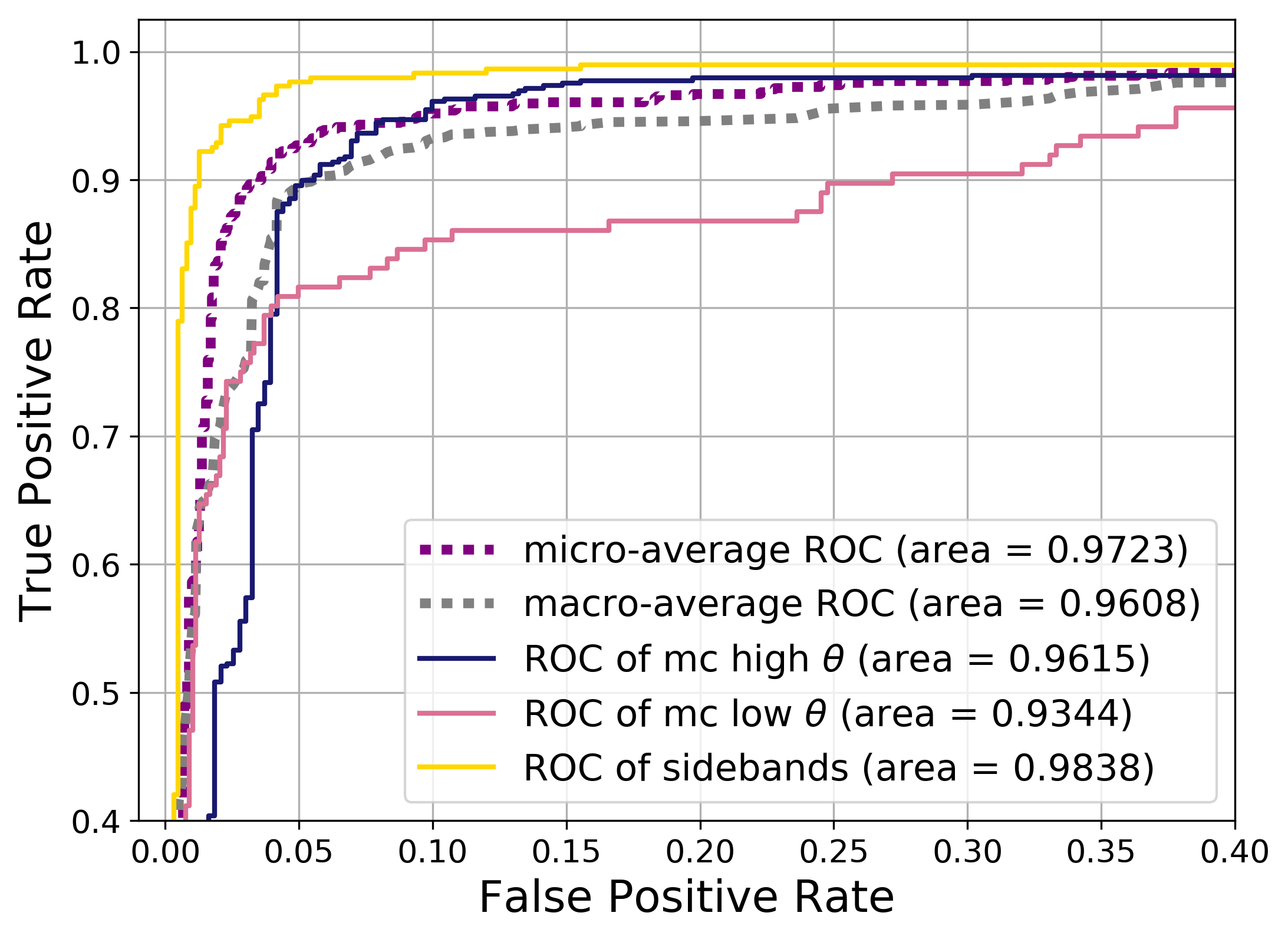}
\caption{ROCs of individual classes and averages for the narrowband model.}
\label{fig:narrowband_roc}
\end{center}
\end{figure}

The respective test set accuracy on the 30 keV range is $91.2\%$. We also generate the ROC curves for each class and average them as shown in Figure \ref{fig:narrowband_roc}. Across all classes we observe an AUROC over 0.9 which indicates that our model does very well at separating any individual type of track from the rest. The ROC curve for low pitch angle mainbands (in pink) has the lowest AUROC, which is most likely due to its relatively small population relative to the other two classes; this comes into effect at One-vs-Rest level where the former population is pitted against the latter simultaneously. Both average curves achieve a ROC over 0.960 putting us in an excellent range of model stability to compliment the high test accuracy obtained.

\begin{figure*}
  \centering
  \begin{tabular}{@{}c@{}}
  \small (a) 30 keV classified frequency spectrum \\
  \includegraphics[width=.75\linewidth]{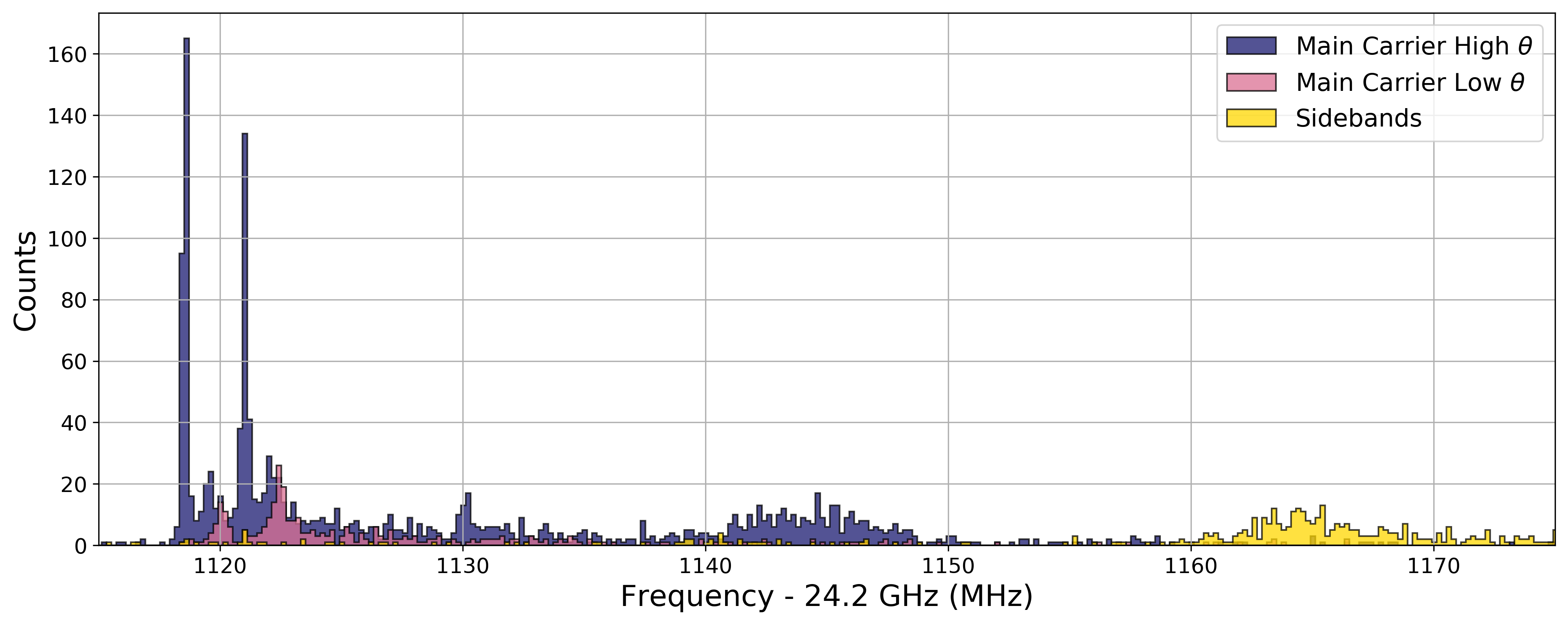} \\[\abovecaptionskip]
  \end{tabular}
  \vspace{\floatsep}
  \begin{tabular}{@{}c@{}}  
  \small (b) 32 keV classified frequency spectrum \\
  \includegraphics[width=.75\linewidth]{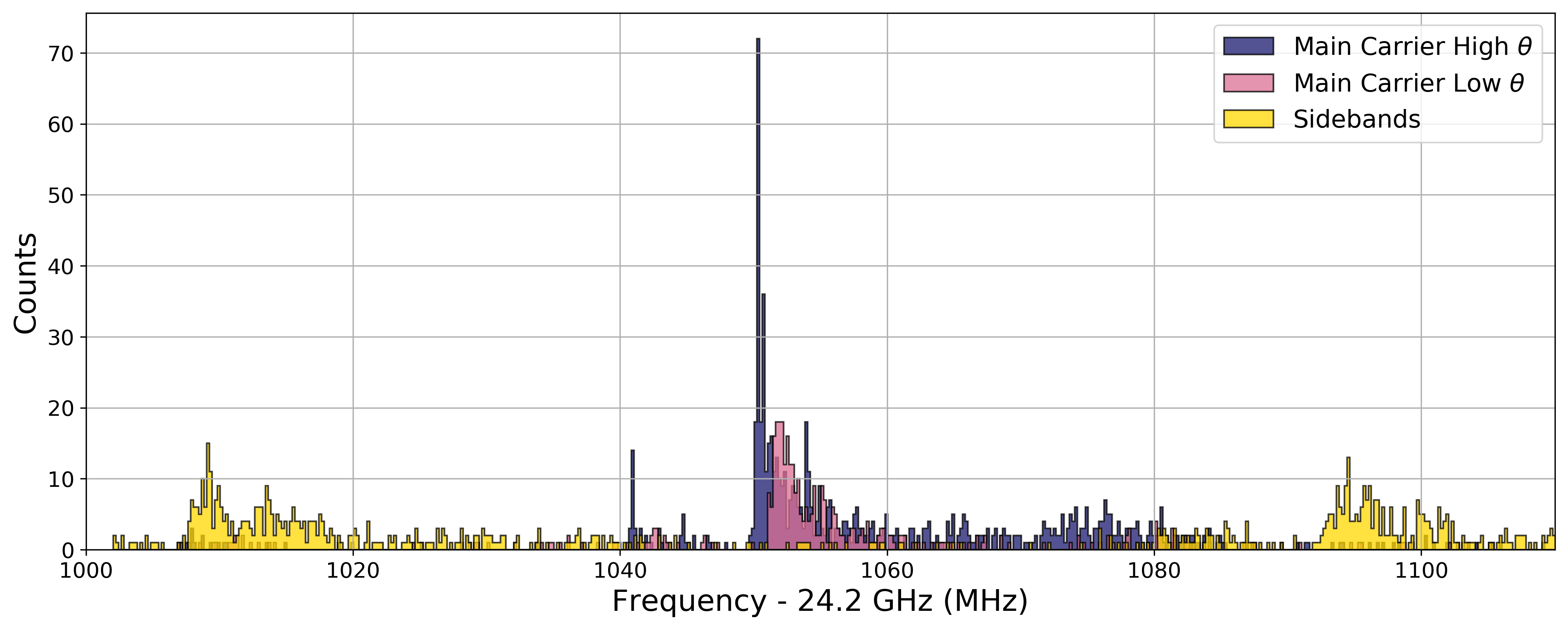} \\[\abovecaptionskip]
  \end{tabular}
  \caption{30 and 32 keV frequency spectra classified with the narrowband model. The colors represent the SVM class identification.}\label{fig:narrowband_spectra}
\end{figure*}

\begin{figure*}
  \centering
  \begin{tabular}{@{}c@{}}
  \small (a) 30 keV slope-frequency correlation \\
\includegraphics[width=.65\linewidth]{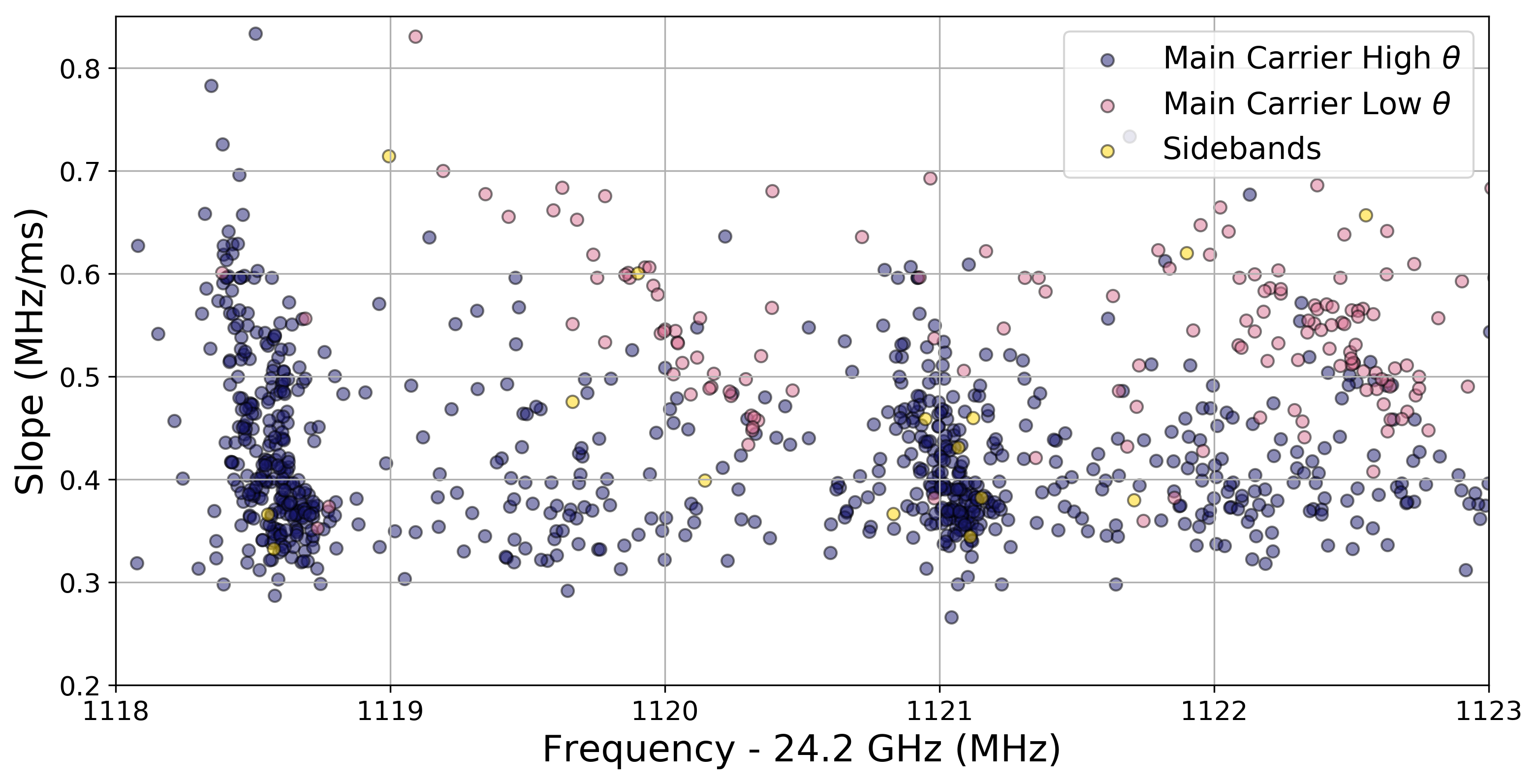} \\[\abovecaptionskip]
  \end{tabular}
  \vspace{\floatsep}
  \begin{tabular}{@{}c@{}} 
  \small (b) 32 keV slope-frequency correlation \\
  \includegraphics[width=.65\linewidth]{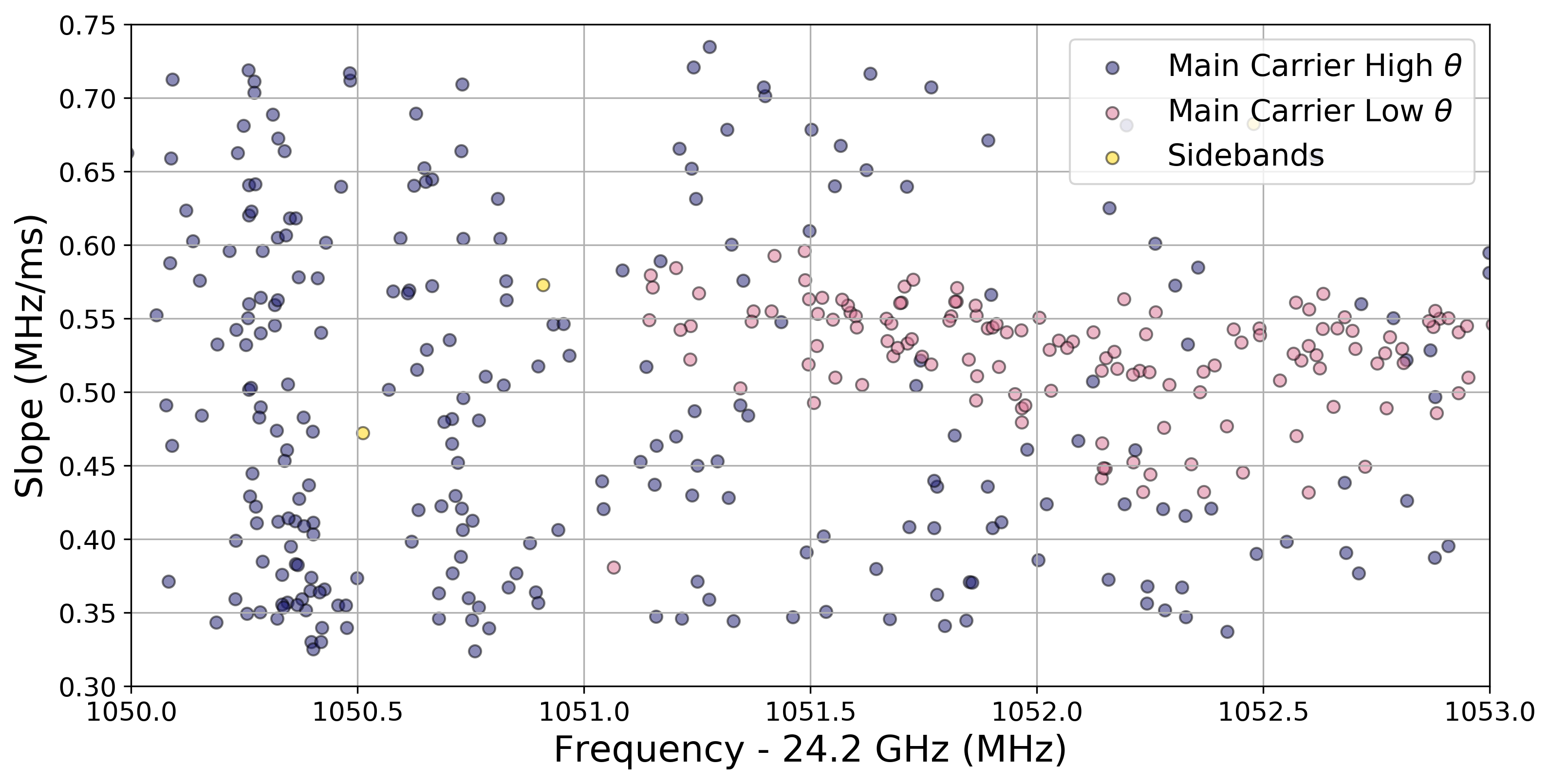} \\[\abovecaptionskip]
  \end{tabular}
  \caption{30 and 32 keV track slope and frequency correlations classified with the narrowband model. The colors represent the SVM class identification.}\label{fig:narrowband_slopes}
\end{figure*}

In Figure \ref{fig:narrowband_spectra}(a) we see the resulting classified track start frequency spectrum in the 30 keV range, the range on which this classifier was trained. We observe a clean separation of mainband tracks in blue (high pitch angle) and pink (low pitch angle) to sidebands in yellow that are mostly concentrated above 1150 MHz. The broad peak between 1160$-$1170 MHz corresponds to upper sidebands of $2^{\textrm{\tiny{nd}}}$ order, giving a rough estimate of the axial frequency $f_a\approx22.5$ MHz; the $n=1$ sidebands are predicted to be suppressed due to the short effect discussed in Section \ref{subsec:cres_signal_waveguide}. To see the separation between mainbands of different types more clearly, we study the slope populations in the frequency range 1118$-$1123 MHz. The overlap of high and low pitch angle mainbands in frequency space seen in Figure \ref{fig:narrowband_spectra}(a) around the low pitch angle peaks is now apportioned, as evident in Figure \ref{fig:narrowband_slopes}(a). The blue scatter points in the region around the low pitch angle peaks constitute true high pitch angle carriers whose true start frequency has been missed during primary track reconstruction. This separation allows a single-valued reconstruction of the pitch angle for a mainband of a given slope; recall that energy corrections may be performed once the pitch angle information is available. 

We also apply the narrowband classifier to the 32 keV group and obtain an accuracy of $92.8\%$, which surpasses the test set score. As can be seen in Figure \ref{fig:narrowband_spectra}(b), the frequency spectrum sports a clean separation between mainbands and sidebands; in this case both upper (around 1090 MHz) and lower (around 1010 MHz) sidebands are visible. The short peak around 1040 MHz has also been classified as a mainband with its respective low pitch angle tail. This corresponds to the 32.14 keV krypton line, which would be extremely difficult to spot by eye, as its relative intensity is very low. The mainbands of the 31.9 keV lines are separated neatly in slope-frequency space as seen in Figure \ref{fig:narrowband_slopes}(b), once again giving way for possible energy reconstruction through pitch angle extraction. The success of the narrowband model at 32 keV is reassuring that a similar technique could be applied to the tritium endpoint region with training on the 17 keV krypton peak.

\subsection{Optimal Set of Classification Features}
\label{subsec:optimal-feats}
We can further improve the performance of the narrowband SVM model by examining how useful each parameter is for accurate classification. An exhaustive search of the unique combinations of features from the 14-dimensional feature space results in $2^{14}-1=16,383$ iterations of the training algorithm. We train a SVM and evaluate the accuracy and the AUROC for each feature subset. The subset with the best overall performance may have improved accuracy compared to the full feature set, and will require less intense computation resources to train on a large data campaign.

To evalulate each subset SVM with a single metric, we use the sum in quadrature of the accuracy and the AUROC value:
\begin{equation}
\Delta_\mathrm{opt} = \sqrt{x^2+y^2}
\end{equation}

\noindent where $x$ is the accuracy of the model and $y$ is the AUROC. The maximum possible value of $\Delta_\mathrm{opt}$ is $\sqrt{2} \approx 1.414$. Maximizing this combined metric allows us to asses both the model stability and its predictive power simultaneously.

Perhaps unsurprisingly, the value of $\Delta_\mathrm{opt}$ is in general large for subsets with many features; a complex model has a greater capability for increased performance. However, some single-feature models are able to achieve high values of $\Delta_\mathrm{opt}$ as well. \texttt{Average} exemplifies this, which alone yields a model with $90.1\%$ accuracy and $\Delta_\mathrm{opt}=1.314$. Of course, models with a single feature for classification run the risk of lacking enough variance to generalize effectively so we discard them as viable candidates. A global maximum of $\Delta_\mathrm{opt}=1.359$ is achieved with a model utilizing 6 final parameters: \texttt{TotalPowerDensity}, \texttt{TrackSlope}, \texttt{TimeLength}, \texttt{MeanCentral}, \texttt{NormCentral}, and \texttt{MaximumCentral}. This amounts to an accuracy of $94.9\%$, an increase of $3.7\%$ compared to the original result in Section \ref{subsec:narrowband_results}, and an AUROC of $0.97$ on the test set.

\subsection{17 keV Peak: Sidebands and Energy Dependence}

The 17 keV peak presents a unique problem for the classifier; as discussed in Section \ref{subsec:cres_signal_waveguide}, the CRES signal power distribution across the sideband spectrum is energy-dependent as a consequence of the waveguide short. We have come to understand that the 17 keV Phase I data studied here consists almost entirely\footnote{A small fraction ($\sim$ 1\%) of observed tracks at 17 keV are hypothesized to be genuine mainband signals from shake-off electrons.} of pairs of $1^{\textrm{st}}$-order sideband tracks, due to a large suppression of the mainband peak from the short interference effect. We have studied more closely the region between the 17 keV sideband peaks and found an excess of power corresponding to an average SNR of 1.26; much too low for track reconstruction, but adequate in combination with other studies to confirm the sideband hypothesis. Consequently, we cannot train the classifier on 17 keV tracks since we have no mainband tracks to offer it.

Instead, we can first use the same classifier that was trained on 30 keV to evaluate the 17 keV tracks, and the results are subpar, with an accuracy of $75.5\%$. While a sizeable portion of tracks are still classified properly, we are confident that nearly all of those classified as main carriers are incorrect. However, this result isn't unexpected; the power and slope correlation discussed in Section \ref{subsec:primary-track-params} is also energy-dependent, and from Section \ref{subsec:optimal-feats} we now understand these two parameters to be among the most decisive in the classification scheme. In the training scheme, the classifier becomes familiar with the 30 keV power-slope correlation, and this has a considerable negative influence when applied to the 17 keV peak where the true power-slope correlation is different. It should be noted that this effect is also present at 32 keV when trained on 30 keV, but from the comparable accuracy scores and AUROCs we conclude it is insignificant for this small energy difference, much to our advantage.

\subsection{Wideband Classifier}

To improve the classifier performance on 17 keV data, we will consider two approaches: first, we use only the rotate-and-project parameters, which have no energy dependence. Second, we train and evaluate the 14-dimensional classifier simultaneously on all three peaks. 

With the rotate-and-project parameters only, we re-train and test the classifier on 30 keV data, and apply it at 17 keV. In the test set (30 keV), the total accuracy decreases to $86.1\%$; this is reasonable given that the slope and power information, which was especially useful in discriminating between the two mainband types, is missing. When applied at 17 keV, we observe an accuracy of $78.9\%$, which is a modest improvement over the narrowband model ($75.5\%$). The ratio of the 17 keV accuracy to that of 30 keV is more substantially improved: 0.917 compared with 0.829 for the narrowband. This suggests, as we hypothesized, that energy-dependent parameters are partially responsible for the shortcoming of the narrowband classifier at 17 keV. However, the rotate-and-project only performance is still far less than ideal and poor compared to the narrowband model at 30 and 32 keV, thus it does not provide us with a satisfactory alternative.

\begin{figure*}[!htbp]
  \centering \includegraphics[width=.75\linewidth]{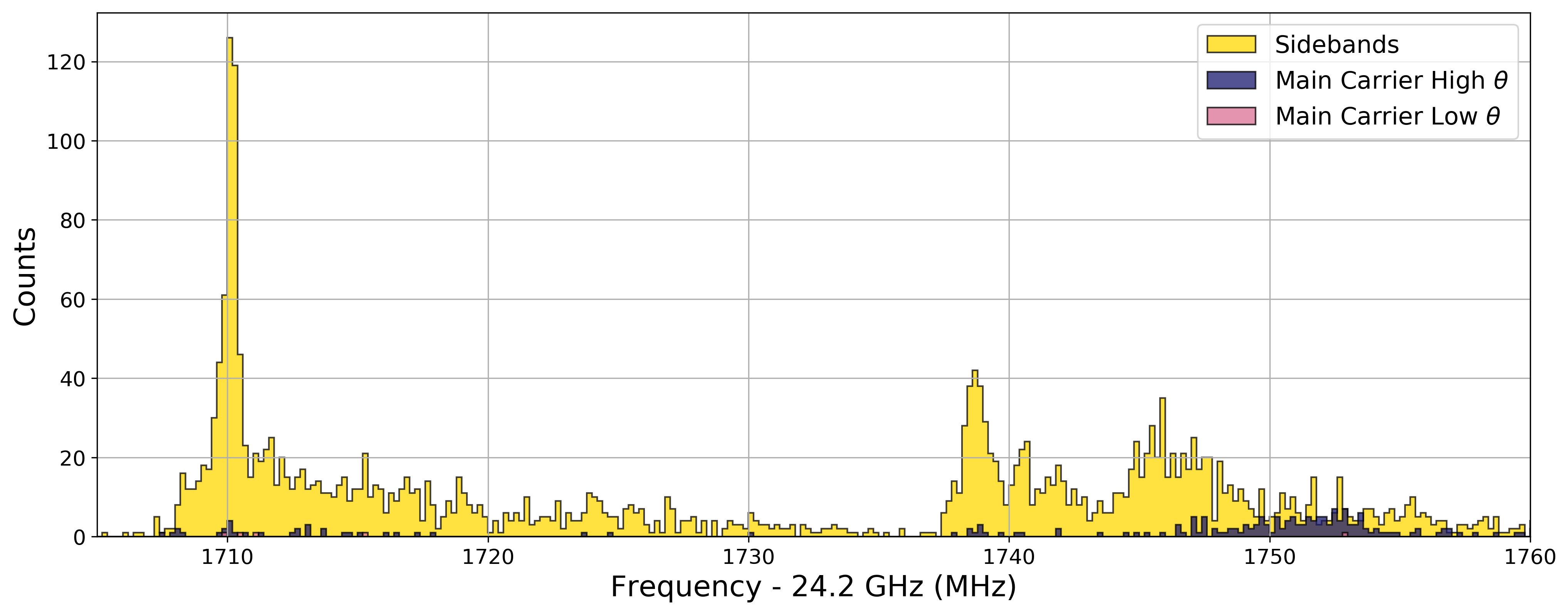}
  \caption{17 keV classified frequency spectrum with wideband model. The colors represent the SVM class identification.}
\label{fig:17keV_spectrum}
\end{figure*}

The second approach we explore considers all three energy ranges simultaneously for training; we call this model the wideband classifier. The 17 keV classified spectrum from this model is shown in Figure \ref{fig:17keV_spectrum}. It is immediately clear that this model has by far the best performance at 17 keV, with an accuracy of $96.1\%$. The largest population of mainband tracks is broadly centered about $\sim$ 1750 MHz, which is at least 30 MHz above the (suppressed) mainband peak. Informal inspection of the slope-power correlation among these tracks, as well as the individual spectrograms, suggests they are indeed most consistent with true mainband tracks. We interpret this as evidence for satellite shake-up/shake-off electrons \cite{shakeup} to be further investigated. Similar broad peaks of mainbands at the 30 and 32 keV ranges have also been observed at a similar separation in frequency (see Figure \ref{fig:narrowband_spectra}). Overall, the 30 and 32 keV lines themselves also show improved accuracy scores compared to the original narrowband model with the full feature set: $92.3\%\ (+1.1\%)$ and $95.6\%\ (+2.8\%)$ respectively.

\section{Discussion and Outlook}

We now have two candidate classifier models: the optimal-feature set narrowband model and the wideband model. Since the classifier will be used for future tritium analyses, we keep the context of tritium data in mind when discussing the advantages of each. The overall accuracy scores and AUROCs for each classifier are summarized in Table \ref{table:model-comparison}\footnote{The training computation times for all three models are on the same order, between 15 to 30 minutes and does not require high performance computing.}.

\begin{table}[h]
\begin{indented}
\lineup
\item[]\begin{tabular}{@{}*{5}{l}}
\br
\0\0Model & \multicolumn{3}{c}{\begin{tabular}{ccc}\multicolumn{3}{c}{Accuracy} \\ 17 keV & 30 keV & 32 keV \end{tabular}} & AUROC\cr 
\mr
Narrowband & \hspace{0.25cm}75.5\% & \hspace{0.15cm}91.2\% & \hspace{0.1cm}92.8\% & \hspace{0.25cm}0.967 \cr
Wideband & \hspace{0.25cm}96.1\% & \hspace{0.15cm}92.3\% & \hspace{0.1cm}95.6\% & \hspace{0.25cm}0.984 \cr
Optimized Narrowband & \hspace{0.45cm}--- & \hspace{0.15cm}94.9\% & \hspace{0.1cm}94.0\% & \hspace{0.25cm}0.973 \cr
\br
\end{tabular}
\end{indented}
\caption{Summary of classification model accuracy scores and averaged AUROC metrics. We exclude 17 keV from the calculation of the narrowband AUROC.}
\label{table:model-comparison}
\end{table}

In the previous section, we improved the classification accuracy at 17 keV by training on all three peaks simultaneously (the wideband model); this model also boasted percent-level improvements to the accuracy at 30 and 32 keV. We understand that the narrowband classifier performed poorly at 17 keV due to energy-dependent correlations between parameters, and the lack of appropriate training at 17 keV. In a tritium analysis, the data acquisition will be contained to not more than $\pm 2$ keV around the endpoint; this includes the 17 keV krypton peak, which can be used for magnetic field calibration. Therefore, if we construct a narrowband model in this context $-$ trained on 17 keV krypton data and applied to tritium data from approximately 15$-$19 keV $-$ it is reasonable to expect a performance similar to the current narrowband model at 30 and 32 keV. Although the narrowband classifier has not been directly trained and evaluated in the tritium endpoint energy range, we have high confidence in its applicability there in the future. The overall results of the various classification approaches and studies have yielded great improvements to our understanding of the data to give us this confidence.

We also saw in the previous section that the choice of an optimal subset of the classification features improved the narrowband model accuracy and AUROC metrics at the percent level, bringing them to a point comparable to the wideband model. In the next section, we will also examine the effect of imperfect classification on the resultant tritium spectrum. Upon comparison of the performance of a narrowband 30 keV network and the wideband network applied to 30 keV and 32 keV data, we see that the wideband model gives only marginal improvement in performance over the narrowband model trained on a close neighbor. Since the tritium signal spectrum has a close neighbor calibration source at 17 keV of Krypton, it is not necessary to bring in the complexity of a wideband model. The sideband problems which prevented a version of the narrowband model from being trained at 17 keV in this work are expected to be ameliorated in hardware at later phases. However, in the eventuality that they remain a challenge, we have shown that a wideband model can achieve good performance as well if it is trained properly.

\subsection{Future Applications in Event Reconstruction}

The signal from a single electron in general takes the form of many reconstructed tracks, via (a) sideband power deposition and (b) scattering interactions with residual gas molecules. The sideband comb structure creates a group of parallel tracks as discussed in Section \ref{subsec:cres_signal_waveguide}, and discrete energy loss from a scattering interaction creates a "jump" in the signal frequencies and in the pitch angle. After track finding, those tracks that belong to the same event (electron) are grouped together to obtain only the start frequency of the event as a whole. In the current event building scheme, this is accomplished with two stages in sequence corresponding to the two items above. First, individual tracks are combined into MPT objects based on a coincidence between start and end times. Many such MPTs are then joined into a single event using a similar coincidence check on the timestamps, this time a head-tail comparison. A full treatment of this process is given in \cite{BenLThesis}. However, the present event builder (with no classifier information) makes no statement about the identity of the mainband within a MPT structure; the start frequency of an event is simply defined as that of the first track (in time) within the first MPT of the event sequence. The classification scheme thus creates the potential for a more intelligent event building procedure which takes advantage of the labeled topologies to determine the true main carrier start frequency of an event.

One simple improvement is to utilize the classification labels to add consistency checks in event building. A MPT structure should logically contain no more than one mainband, and the start frequency of the MPT should be determined by this mainband track alone (if present). MPTs with two or more sidebands may also be checked to ensure the frequency spacing between them is consistent with a unique axial frequency. If a mainband track decreases in frequency after a scatter, indicating a sharp increase in the pitch angle, the accompanying decrease in axial frequency of the candidate sidebands may also be used as a check. These examples are only some of the many possibilities in which classification labels can enhance our reconstruction process. 

\begin{figure*}[ht!]
\centering
\includegraphics[width=0.95\linewidth]{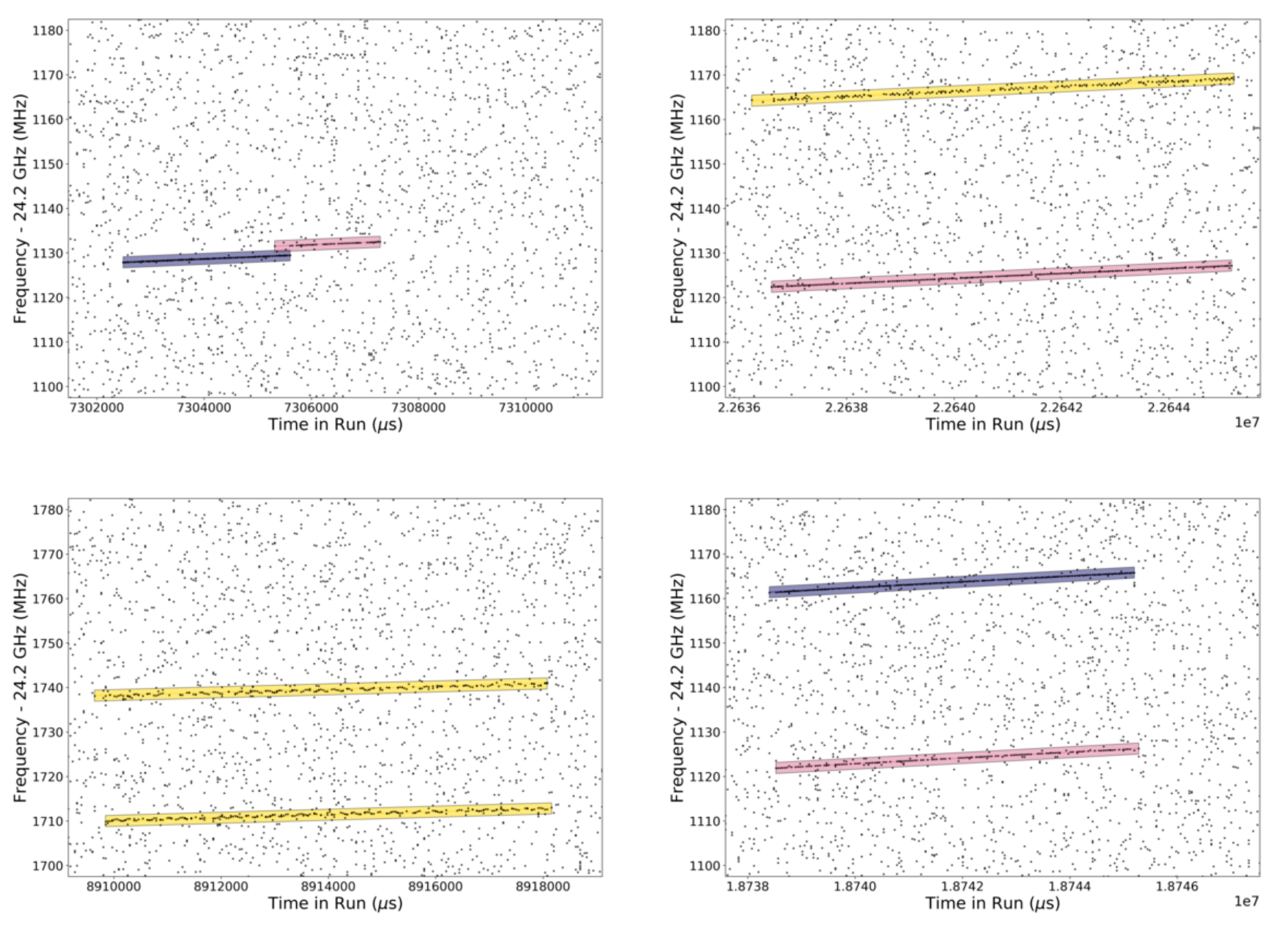}
\caption{Classified tracks of candidate MPT events exhibiting multiple topological combinations present in Project 8 Phase I data. The blue are mainband high pitch angle, the pink mainband low pitch angle and the yellow sideband classified tracks. The rectangular boxes are for illustration only; the track is composed of all points concentrated along a line passing through the middle of the box.}\label{fig:event_multi}
\end{figure*}

\begin{figure}[ht!]
\centering
\includegraphics[width=0.85\linewidth]{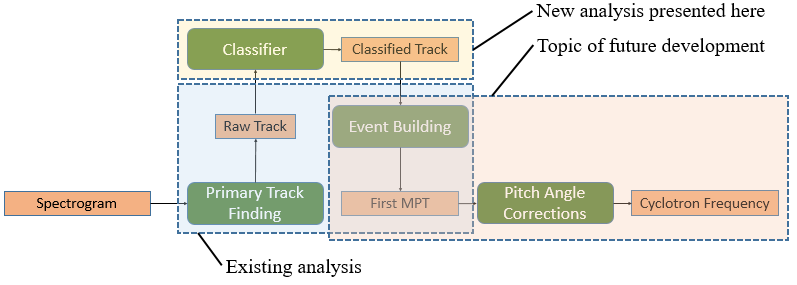}
\caption{Analysis flowchart described and proposed in this work. The green blocks indicate large-scale processing steps and the smaller orange blocks show the data at each step. Feature extraction and the classifier decision function are contained within the block labelled `Classifier', and the classified track provides more input information to the event builder compared to the raw track. Expansion of the event building stage and implementation of the pitch angle corrections are the most critical future analysis tasks to fully utilize the classification scheme.}
\label{fig:flowchart}
\end{figure}

In Figure \ref{fig:event_multi} we show some typical interesting event topologies which could benefit from an event builder that utilizes the classifier results:

\begin{itemize}
    \item {Top left: An event with a scatter that changes the pitch angle from the high to low region, according to the classifier. A faint sideband is also visible above the mainband after the scatter, but it was not reconstructed. It is interesting to note that this change in topology would be very difficult for a human labeller to identify, but is clearly easy for the classifier.}
    \item{Bottom left: A MPT with two lone sidebands. In this case, we can reconstruct the hidden mainband start frequency from the axial separation and determine the pitch angle correction.}
    \item{Bottom right: A MPT that the classifier has identified to contain two mainbands. This indicates an error, either in the classification or the MPT construction. To address this, we might consider the relative probabilities that either track is in fact a sideband (i.e. work event building information into the classifier), or simply discard the event if it cannot be made sensible.}
\end{itemize}
An improved event builder that works in tandem with the classifier is crucial for proper reconstruction of a continuous tritium spectrum using CRES. Complex event topologies and sideband proliferation from lower energies in the spectrum continuum will demand a sophisticated understanding of the underlying nature of tracks. For example with atomic tritium, assuming $1\times 10^{18}$ atoms/$\textrm{m}^3$ and a cylindrical voxel\footnote{Assuming that we have beamforming in the radial direction with 1 cm or better resolution, and no axial information.} of 1 cm in diameter and 10 m in length, for events with, on average, ten tracks of length \SI{80}{\micro\second} each we expect $3.14\times 10^{15}$ atoms/voxel. Then, given an activity of \SI{5.6e6}{\becquerel} per voxel for the entire spectrum, we expect about 1.69 events (about two events) present at all times when looking at a 1 keV window below the endpoint where only a $2\times 10^{-4}$ fraction of the activity is present. With two possibly overlapping events in a given spectrogram, the use of an accurate classifier will be decisive in identifying and separating the constituents of each; we expect that the model presented here is a decisive step toward that success. In Figure \ref{fig:flowchart} we outline the analysis steps discussed for this work, including the future work regarding (a) the event builder as discussed in this section, and (b) pitch angle corrections to extract the true cyclotron frequency.

\subsection{Effects of Misclassification on Tritium Spectrum}
\label{subsec:effect_on_tritium}

Along with the improved event builder, a classifier helps us reuse or remove all but the misclassified sidebands at a high confidence level. However, it is still important to study the effect of sideband proliferation through misclassification on the tritium spectrum. Using the Morpho \cite{morpho} interface to perform Hamiltonian Monte Carlo simulations with the Stan package \cite{stan}, we model the electron kinematic variables and compute the detected track frequencies according to the discussions in Section \ref{subsec:cres_signal_waveguide}. The kinetic energy is drawn from the tritium spectrum probability distribution function with an endpoint of exactly $Q=18600$ eV and zero neutrino mass. The power in each of the mainband and the pair of $n=2$ sideband tracks is then calculated for a circular waveguide with the same bathtub trap configuration as for the data used in this paper. For those electrons that become trapped, a uniform detection threshold is enforced on each track. The detected tracks are collected into a mainband spectrum, denoted $p_0(E)$, and a sideband spectrum denoted $p_2(E)$\footnote{Recall that the $n=1$ sidebands are suppresed due to the interference effect.}. Here, $E$ represents the inferred kinetic energy from the detected track frequency, which in the case of sideband tracks would constitute an erroneous reconstruction; this allows us to study the effects of both misclassification and the lack of pitch angle corrections. It is important to note that this simulation serves only as a toy model and does not reflect the complete design of the Project 8 detector in Phase I (from which the data presented here in earlier sections was taken), or Phase II which has since reconstructed the first ever tritium spectrum with CRES. However, the toy model is relevant as a mean to highlight and evaluate some of the challenges that sideband presence will bring when reconstructing any CRES spectrum. 

If the probability for a track to be wrongly classified (either mainband as sideband, or vice-versa) is uniformly $\alpha$, then the spectrum of classified mainband tracks is:
\begin{equation}
p(E)=(1-\alpha)p_0(E)+\alpha\ p_2(E)
\end{equation}

\begin{figure}[h!]
\begin{center}
\includegraphics[width=0.6\textwidth]{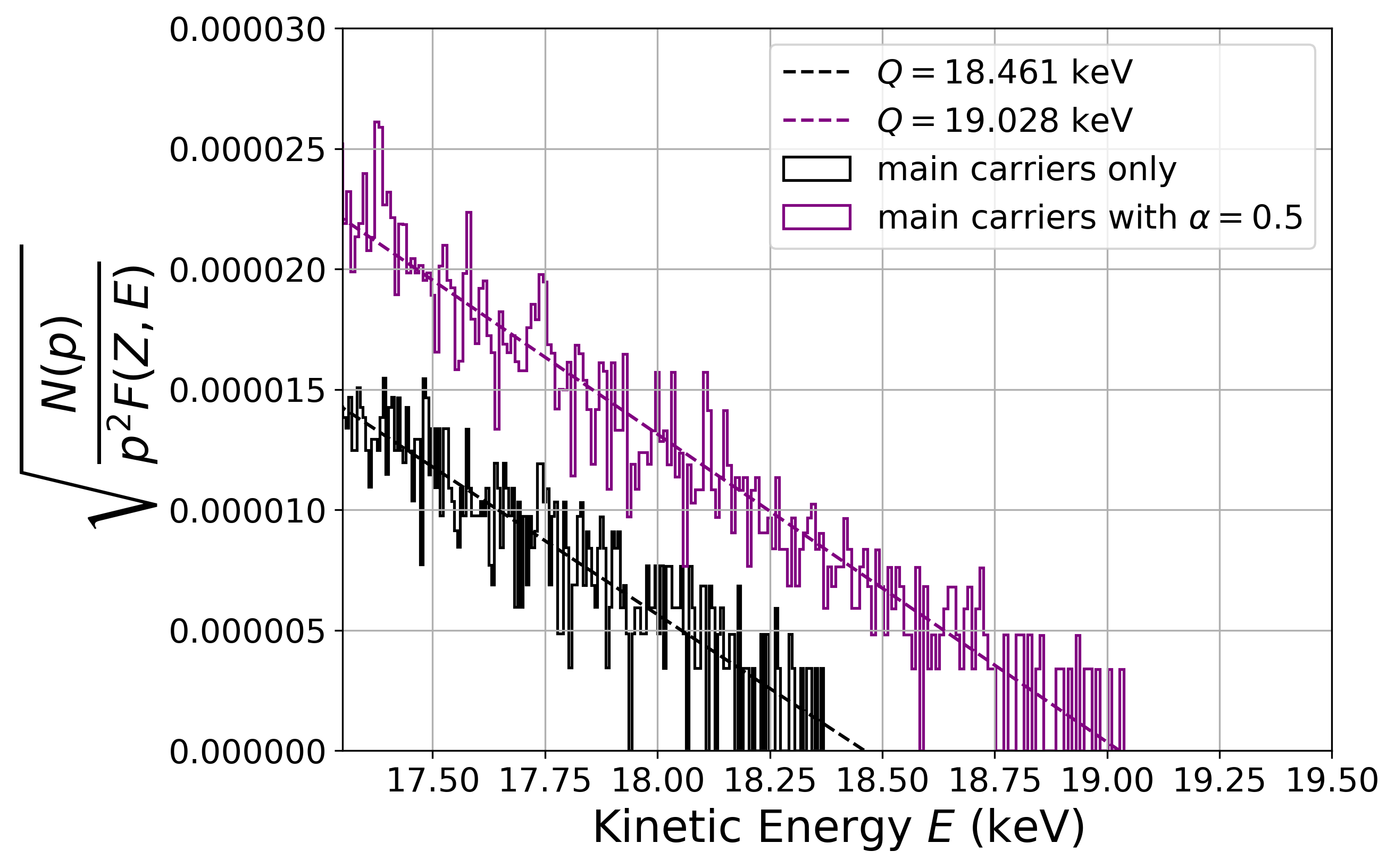} 
\caption{Kurie plot of two simulated tritium spectra in the toy model described in section \ref{subsec:effect_on_tritium}: mainbands only (black) and sideband-contaminated with misclassification rate $\alpha=0.5$ (purple).}\label{fig:kurie_plot}
\end{center}
\end{figure}

\noindent With $\alpha=0$, we obtain the main-carrier-only spectrum from a "perfect" classifier. Since the energy of each event is calculated from the measured (mainband) frequency and not the cyclotron frequency, we expect to observe a shift to lower energy (higher frequency) in the endpoint. Indeed, Figure \ref{fig:kurie_plot} shows the simulated Kurie plot for perfect classification $\alpha=0$ (in black) and the fitted $Q$-value:
\begin{equation}
Q(\alpha=0)=18.461\ \mathrm{keV}
\end{equation}
which deviates from the true input value by 139 eV. In the same figure we show the spectrum with $\alpha=0.5$ (random classification); now, the region above the endpoint is contaminated by sidebands. As a result, the endpoint is measured at:
\begin{equation*}
Q(\alpha=0.5)=19.028\ \mathrm{keV}
\end{equation*}
which exceeds the true value by 428 eV, or roughly 20 MHz: quite similar to the observed axial frequency.

\begin{figure}
\begin{center}
\includegraphics[width=0.5\textwidth]{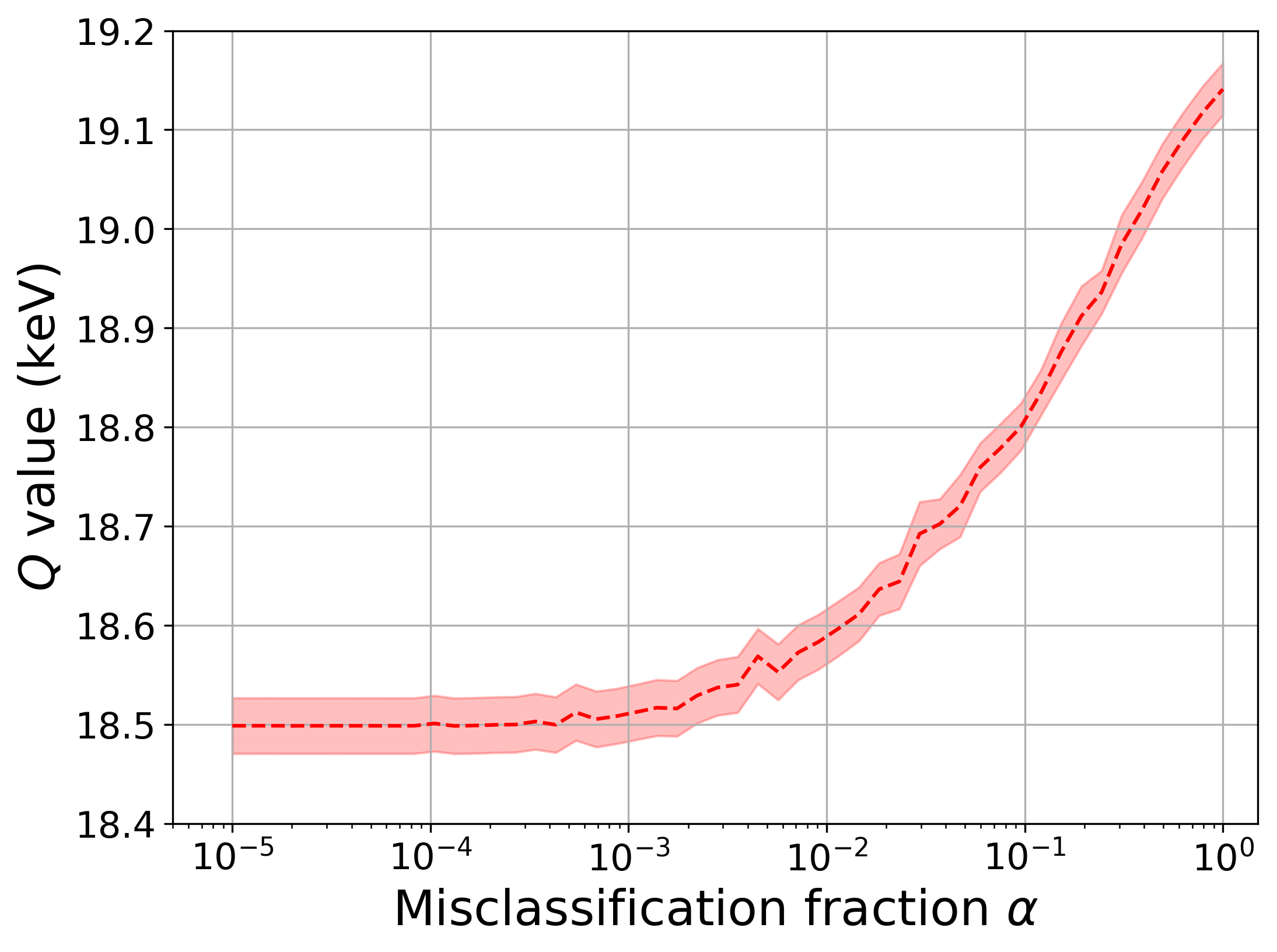}
\caption{Extracted $Q$-value from Kurie fit to sideband-contaminated tritium spectrum simulations for different values of misclassification fraction $\alpha$. The light red band represents the standard error of the mean. The true value of the endpoint is asymptotically approached for decreasing $\alpha$ which, without energy corrections, sits around 18.5 keV.}\label{fig:q_vs_alpha}
\end{center}
\end{figure}

Lastly, we perform the Kurie fit for many values of $\alpha$ to see the dependence of $Q$ in a more continuous form; Figure \ref{fig:q_vs_alpha} illustrates these results. Each $Q$-value (shown as the dashed line) represents the mean of 50 unique spectrum simulations for a fixed value of $\alpha$; the band in light red spans the standard error of the mean on either side, and is dominated by the statistical uncertainty of each simulation. Though the simulations are themselves independent, we use the same set of simulations for all values of $\alpha$. Thus, the uncertainty band in $Q$ does not reflect the endpoint measurement uncertainty of the simulation, nor, more importantly, of any real Project 8 phase. Comparing to $\alpha=0$, we observe that even for $\alpha \approx 10^{-2}$ which is small compared to our demonstrated models, the endpoint shift is significant; a precise requirement or bound on $\alpha$, however, is again not necessarily transferable to Project 8. Still, we may safely conclude that going forward with the classification models we must have the highest reasonable standard for accuracy. Our thorough understanding of the trap geometry and related systematics helps in addition to reduce the effects of sideband contamination in future Project 8 phases, through both design and trap configuration.

We have clearly demonstrated with this simulation that both energy corrections and track classification will have a substantial influence on future CRES tritium results, with an observed endpoint shift on the scale of $\mathcal{O}$(100 eV). The design and configuration of future phases of Project 8 are guided in part by the goal to suppress detectable sidebands and achieve sub-percent level misclassification. Incorporation of this improved track classification and pitch angle considerations will enable a CRES experiment like Project 8 to take the next step and achieve a competitive eV-scale endpoint sensitivity.

\section{Conclusions}

With the phenomenological model put forward in \cite{PhenomPaper}, we have motivated the need for a classification of CRES signal topologies according to the pitch angle distribution of trapped electrons and the sideband comb structure of events discussed in Section \ref{sec:need_for_classification}. With the use of reconstructed signal properties including total power, track slope, and analysis of the rotated-projected spectrum (Section \ref{subsec:rotate-project}), we have enumerated 14 quantities that have the power to discriminate between the three different track topologies. We have implemented a Machine Learning-based classification scheme using a Support Vector Machine algorithm with these 14 parameters, and studied the results of several models on Project 8 Phase I data. With the use of an optimal feature set, we have achieved a model with $94.9\%$ total accuracy and an AUROC of $0.973$. This model is trained and applied over a total energy range of approximately 3.5 keV, which gives us confidence in its applicability to tritium analysis ($\sim 1$5$-$19 keV).

The use of these classification models has already improved our understanding of the Phase I data. It has bolstered our confidence in the phenomenological model, our understanding of the relationship between signal characteristics and trap geometry, and the set of optimal features provides some insight into the nature of tracks both physically and from the viewpoint of reconstruction. The classified spectra are much more informative than ordinary (unclassified) spectra, and our comparison of different training models has highlighted the energy dependence of some track parameters.

Energy correction $-$ from measured frequencies to true cyclotron frequencies $-$ is another necessary step that will utilize track classification, but this application is outside the scope of this paper. These corrections were discussed briefly in Section \ref{subsec:energy_corrections}, and in Section \ref{subsec:effect_on_tritium} we showed through simulation that we expect these corrections to be of order 100 eV. Consequently, for an experiment like Project 8 to perform an eV-scale precision measurement of the tritium spectrum in its next phase, it is essential to minimize the impact of sideband effects. Strategies to accomplish this have been pursued already in Phase II and are a key consideration in the development of the Phase III experiment; one such strategy in Phase II almost completely eliminates the presence of detectable sidebands by reducing the frequency modulation index. We also discussed future prospects of an event builder that works in tandem with the classifier, and perhaps through Machine Learning as well. With these types of improvements to the apparatus guided by this work, and a highly robust track/event classification scheme, Project 8 will work toward a Phase III analysis that is greatly advanced and mature compared to earlier phases, and capable of achieving an endpoint measurement with eV-scale precision.

Project 8 has demonstrated the CRES technique, constructed the first-ever CRES tritium spectrum with it, and will soon be looking toward a competitive eV-scale measurement of the neutrino mass limit in Phase III. The work presented here has contributed greatly to our understanding of CRES signals and the obstacles to an eV-scale sensitivity, and its conclusions have provided us with valuable knowledge of the path toward a highly sensitive measurement. By utilizing the full potential of track classification, we can continue to advance the CRES technique and make valuable steps towards the future ambitions of Project 8: ultra-precise spectroscopy, a decisive measurement of the tritium endpoint, and and ultimately a direct mass measurement.

\section{Acknowledgements}

This material is based upon work supported by the following sources: the U.S. Department of Energy Office of Science, Office of Nuclear Physics, under Award No.~DE-SC0011091 to the Massachusetts Institute of Technology (MIT), under the Early Career Research Program to Pacific Northwest National Laboratory (PNNL), a multiprogram national laboratory operated by Battelle for the U.S. Department of Energy under Contract No.~DE-AC05-76RL01830, under Early Career Award No.~DE-SC0019088 to Pennsylvania State University, under Award No.~DE-FG02-97ER41020 to the University of Washington, and under Award No.~DE-SC0012654 to Yale University; the National Science Foundation under Award Nos.~1205100 and 1505678 to MIT. This work has been supported by the Cluster of Excellence ``Precision Physics, Fundamental Interactions, and Structure of Matter'' (PRISMA+ EXC 2118/1) funded by the German Research Foundation (DFG) within the German Excellence Strategy (Project ID 39083149); Laboratory Directed Research and Development (LDRD) 18-ERD-028 at Lawrence Livermore National Laboratory (LLNL), prepared by LLNL under Contract DE-AC52-07NA27344; the MIT Wade Fellowship; the LDRD Program at PNNL; the University of Washington Royalty Research Foundation.  A portion of the research was performed using Research Computing at PNNL.  The isotope(s) used in this research were supplied by the United States Department of Energy Office of Science by the Isotope Program in the Office of Nuclear Physics.  We further acknowledge support from Yale University, the PRISMA Cluster of Excellence at the University of Mainz, and the Karlsruhe Institute of Technology (KIT) Center Elementary Particle and Astroparticle Physics (KCETA).

\appendix

\section{Supervised Classification Details}\label{appendix}

\begin{figure}[h!]
\begin{center}
\centering
\includegraphics[width=14cm]{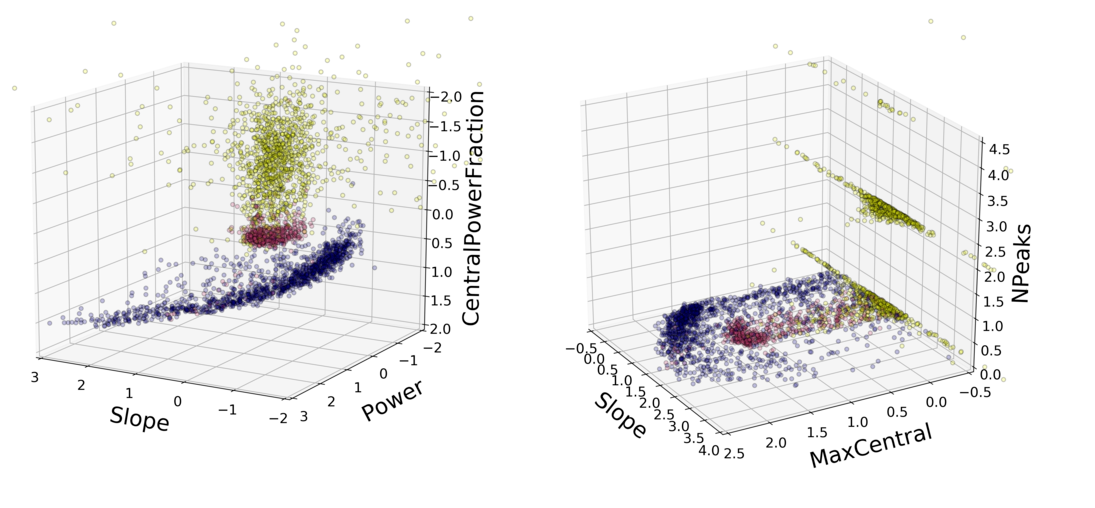}
\caption{3D slices of the classification feature space for the narrowband classifier where the SVM is trained; the class separations can be seen by eye. The blue scatter are mainband high pitch angle, the pink mainband low pitch angle and the yellow sideband tracks. All features are scaled (unitless). \label{fig:training_points}}  
\end{center}
\end{figure}

The Support Vector Machine (SVM) \cite{Cortes1995} is a model that, as all ML algorithms, features a unique loss function to be minimized over a set of $m$ training points $\{x^{(i)},y^{(i)}\}$ with the goal of obtaining a fit parameter vector $w$. The $x^{(i)}$ in our case are electron tracks, each represented as a 14-dimensional vector of features with a label $y^{(i)}\in \{0,1,2\}$: 0 for high pitch angle mainbands, 1 for low pitch angle mainbands, and 2 for sidebands. The optimal fit vector $w$ is also 14-dimensional and may be thought of as defining the normal to a hyperplane in a transformed feature space which maximally separates disjoint classes of points. Classification itself is performed by projecting data points $x^{(i)}$ onto $w$ following constraints regarding the sign of the projection. Specifically, we employ a SVM with a radial basis kernel to facilitate classification of our non-linearly separable track topologies. Example slices of the 14-dimensional space where the SVM trains are shown in Figure \ref{fig:training_points}.

To aid in minimization of the loss function, we rescale all learning points such that each feature has a mean of 0 and a standard deviation of 1 (now in dimensionless units). In a SVM, classification relies on computing a measure of distance from a point to the decision hyperplane. This objective is aided by the scaling, preventing any feature with a widely different range to bias the prediction.

\subsection{Cross-validation and Hyperparameter Optimization}\label{appendix:cv-hyperparam}
The SVM features two hyperparameters, $C$ and $\gamma$, which are, by definition, external configurations that cannot be estimated from data. $C$ encodes the leniency of the SVM in trading misclassification for model stability and $\gamma$ dictates the influence of training points defining the decision boundary to the rest. The quantity $C$ may influence overfitting; a high value gives less weight (penalty) to the higher order $w$ terms in the SVM loss metric allowing them to dominate the minimization strategy. On the other hand, a high $\gamma$ may lead to high bias since the sphere of influence of the support vectors themselves is sharper around the boundary and less so at other points, while a small $\gamma$ usually leads to low variance as the sphere of influence of the support vectors is large. It is thus important to keep this balance in mind during training.

To optimize the hyperparameters we fix two values of $C$ and $\gamma$, fit the SVM to the training set and compute an accuracy score:

\begin{equation}\label{eq:accuracy}
\textrm{accuracy}\left(y,\hat{y}\right)=\frac{1}{m}\sum_{i=0}^{m-1}1\left(\hat{y}^{(i)}=y^{(i)}\right)
\end{equation}
where is $\hat{y}^{(i)}$ the predicted label for point $\{x^{(i)},y^{(i)}\}$ and the optimal model chosen is one for which this score is maximal. In order to avoid overfitting, we perform optimization using a $K$-folds strategy. This scheme involves splitting the original training set into $K$ subsets (folds) and, having picked a $C$ and $\gamma$ tuple, for each fold: 

\begin{enumerate}
\item{Train the SVM using $K-1$ of the folds}
\item{Compute the accuracy score on the remaining part of the data.} 
\end{enumerate}
The subset used in step one is now called the training set and the subset used in step two is named the cross-validation set, both defined dynamically at every step. The partitioning into $K$ folds is random and uses a stratified approach that keeps the relative class ratios in check. After $K$ such iterations, each fold acting as the cross-validation set exactly once, the accuracy scores are averaged and recorded. The hyperparameter domain of $C$ and $\gamma$ itself is defined via a randomized grid search technique for which we have supplied prior exponential distributions to both and set a total iteration value to 100 instances per tuple. The range of values explored were from 0 to 1000 and from 0 to 1 for $C$ and $\gamma$ respectively. The hyperparameter values reported in this work follow from this analysis.

\subsection{Receiver Operating Characteristic Curves}\label{appendix:roc}
\begin{figure}
\begin{center}
\includegraphics[width=0.35\textwidth]{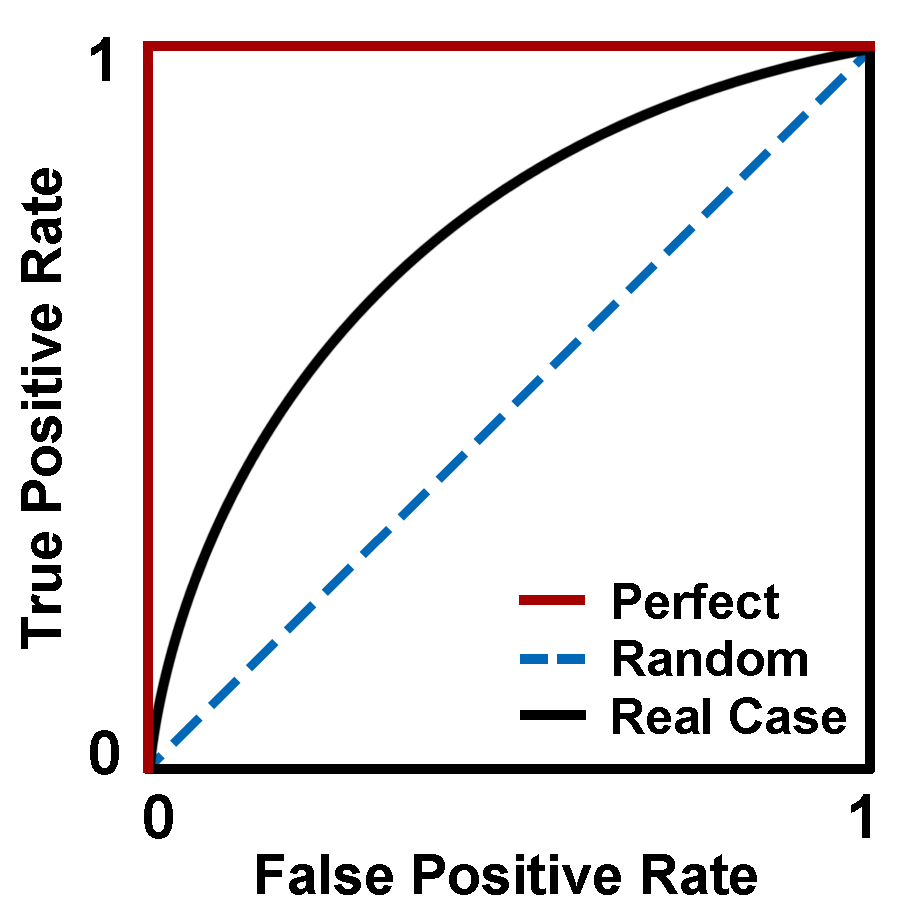}
\caption{Illustration of resulting ROC curves for three cases: a perfect classifier (red) with AUROC 1.0, a random classifier (blue) with AUROC 0.5, and a real-case of intermediate strength (black) following the top figure.}
\label{fig:roc_cartoon}
\end{center}
\end{figure}

Having obtained optimized hyperparameters we move to asses the SVM competency on the test set where the relevant performance metrics are the accuracy (as in Equation \ref{eq:accuracy}) and the Receiver Operating Characteristic curve (ROC). The ROC is a measure of the robustness of the model across all classification thresholds. At every threshold step considered, we compare the rates of true positives and false positives for any classification label to obtain the ROC curve; see Figure \ref{fig:roc_cartoon} for an illustrative example. Given a specific class, a true positive is a track with the correct label and a false positive is a track with any other incorrect label. A random (useless) classifier would produce equal numbers of true and false positives regardless of threshold. With an ideal classifier, every threshold that retains at least one point would yield 100\% true positives. In practice, a classifier ROC curve should lie somewhere between these two scenarios, resulting in an Area Under ROC (AUROC) bounded between 0.5 and 1. The steepness of the ROC curve tells us the power of discrimination and thus the AUROC is a good quantitative measure of the strength of the model itself. 

In order to compute a ROC curve we first binarize the multi-class problem by one hot encoding of the labels. The metrics are then computed using a One-vs-Rest strategy where a single class is pitted against the other two simultaneously. To obtain a comprehensive ROC curve, we may take one or both of the following approaches:
\begin{itemize}
\item{Micro-averaging: count true and false positives in all classes together as a single problem, and then compute the resulting ROC curve}
\item{Macro-averaging: construct the ROC for each classification type separately, then simply average the results.}
\end{itemize}
With the accuracy score, optimized hyperparameters, and ROC curves defined, we evaluate the performance of the classification scheme.

\section*{References}

\bibliographystyle{iopart-num}
\bibliography{biblio}
\end{document}